\documentclass[3p,times, twocolumn]{elsarticle}

\usepackage{amssymb}
\usepackage{amsmath}

\usepackage{dblfloatfix}
\usepackage{geometry}             
\usepackage{graphicx}
\usepackage{epstopdf}
\usepackage{xargs}                 
\usepackage[pdftex,dvipsnames]{xcolor}  
\usepackage[colorinlistoftodos,prependcaption,textsize=tiny]{todonotes}
\usepackage{siunitx}
\usepackage{multirow}
\usepackage{hhline}
\usepackage{caption}
\usepackage{subfig}
\usepackage{verbatim}
\usepackage{verbatimbox}
\usepackage{xcolor, colortbl}

\newcommand{\Figref}[1]{Figure~\ref{#1}}
\newcommand{\tabref}[1]{Table~\ref{#1}}
\newcommand{\Eqref}[1]{Equation~\ref{#1}}
\newcommand{\secref}[1]{Section~\ref{#1}}

\journal{Nuclear Instruments and Methods in Physics Research A}

\begin{document}

\begin{frontmatter}

\title{On the accuracy of Monte Carlo based beam dynamics models for the degrader in proton therapy facilities}

\author{V.\ Rizzoglio\corref{corr1}} 

\author{A.\ Adelmann\corref{cor2}}
\ead{andreas.adelmann@psi.ch} 
\cortext[cor2]{Corresponding author}

\author{C.\ Baumgarten\corref{cor3}}
\author{D.\ Meer\corref{cor4}} 
\author{J.\ Snuverink\corref{cor5}}
\author{V.\ Talanov\corref{cor6}}

\address{Paul Scherrer Institut, 5232 Villigen, Switzerland}


\begin{abstract}
In a cyclotron-based proton therapy facility, the energy changes are performed by means of a degrader of variable thickness.\ The interaction of the proton beam with the degrader creates energy tails and increases the beam emittance.\ A precise model of the degraded beam properties is important not only to better understand the performance of a facility already in operation, but also to support the development of new proton therapy concepts.\ 
The accuracy of the degraded beam properties, in terms of energy spectrum and transverse phase space, is influenced by the approximations in the model of the particle-matter interaction.\ In this work the model of a graphite degrader has been developed with four Monte Carlo codes:\ three conventional Monte Carlo codes (FLUKA, GEANT4 and MCNPX) and the multi-purpose particle tracking code OPAL equipped with a simplified Monte Carlo routine.\ From the comparison between the different codes, we can deduce how the accuracy of the degrader model influences the precision of the beam dynamics model of a possible transport line downstream of the degrader.
\end{abstract} 


\end{frontmatter}


\section{Introduction}
\label{intro}

In particle therapy facilities, the depth-dose distribution to the tumor requires the delivery of different beam energies.\ In a cyclotron-based facility the different energies are obtained slowing the proton beam down in a degrader of variable thickness \cite{HPaganetti}.\ However, a consequence of the degradation process is the increase of the beam emittance and energy spread.\

In the last years, several studies have been performed to improve and optimise the efficiency of the energy degrading process.\ Besides graphite, the use of alternative materials, such as beryllium \cite{Anferov2007} or boron carbide \cite{Gerbershagen2016}, was investigated to minimize emittance growth by energy degradation.\ In the same way, different degrader geometries were proposed to minimize the beam losses and limit the beam phase space \cite{Anferov2003, Lomax}.

These studies are normally performed using fully integrated Monte Carlo (MC) codes (e.g.\ FLUKA \cite{Fluka}, GEANT4 \cite{Geant4}, MCNPX \cite{MCNPX}).\ The energy loss, elastic and inelastic scattering and secondary particle production due to the proton interaction with the degrader can be modelled precisely.\ In other studies, some approximations are used, for example assuming different approaches for the multiple Coulomb scattering \cite{Gottschalk2010}, small angle scattering \cite{Schneider2001} or thin degrader with negligible variation of the particle momentum \cite{Farley2005}.\ In these cases, the model accuracy of the particle-matter interaction is of course reduced in comparison with the results from the general MC codes.

For a cyclotron-based proton therapy facility, a precise particle-matter interaction model for the degrader allows a better understanding of important beam parameters such as the reference energy, transverse emittance and beam current at the degrader exit.\ The undesired side-effects of the degradation process are compensated, at the cost of beam intensity, by the use of a pair of collimators that reduce the beam phase space to match the acceptance of the transport line.\ The energy spread is controlled by means of an energy selection system (ESS), i.e.\ an horizontal slit in the dispersive area between two bending magnets downstream of the degrader.\ The accuracy of the degrader model determines the precision of the predicted proton beam properties along the transport line downstream of the degrader as well as the losses at the collimators and at the ESS \cite{VanGoethem2009}.

Here we investigate how the accuracy of the degrader model is influenced by the use of different particle-matter interaction algorithms and approximations.\ In particular, we compare the results of a graphite degrader simulated using with three fully integrated conventional MC codes (FLUKA, GEANT4 and MCNPX) and with a multi-particle accelerator tracker, called OPAL, equipped with a simplified MC model for particle-matter interaction \cite{opal:1}.\ In order to obtain precise and reliable predictions of the beam properties, the use a tracking code with the ability to perform MC simulations of particle-matter interaction is of advantage.\ The multi-purpose particle tracking code OPAL has this capability and its potential in such an application has been already proven in \cite{Rizzoglio2017}.\ 

The comparison between the four MC codes is performed on the main beam parameters (e.g.\ degraded energy spectrum, growth of the phase space volume, contribution of the inelastic scattering to the total spectrum) which are normally used as starting conditions to develop the beam dynamics model of the transport line downstream of the degrader.\ Our goal is to deduce how the accuracy of a certain particle-matter interaction model influences the beam parameters after the degrader and hence the precision of the beam dynamics model of the subsequent transport line.

In \secref{sec:model}, the model setup used in this work is described.\ The main features of the four MC codes are summarized in \secref{sec:codes}.\ The methods developed for the analysis and results of the comparison are presented in \secref{sec:results}.\ In \secref{sec:Application} the influence that the degraded beam parameters from the four MC codes have on the model of the transport line downstream of the degrader is discussed.

\section{The model setup}
\label{sec:model}

The model described in this work is based on the graphite degrader installed in the PROSCAN facility at the Paul Scherrer Institut (PSI) in Switzerland \cite{PROSCAN}.\ In this facility, a 250~MeV proton beam is extracted from the superconducting cyclotron COMET and focused by a quadrupole triplet onto the degrader, which consists of two pairs of three movable graphite wedges (see \secref{subsec:degrader} for more details). 

\begin{figure}[h!]
  \centering
  \includegraphics[scale=0.11, keepaspectratio=true]{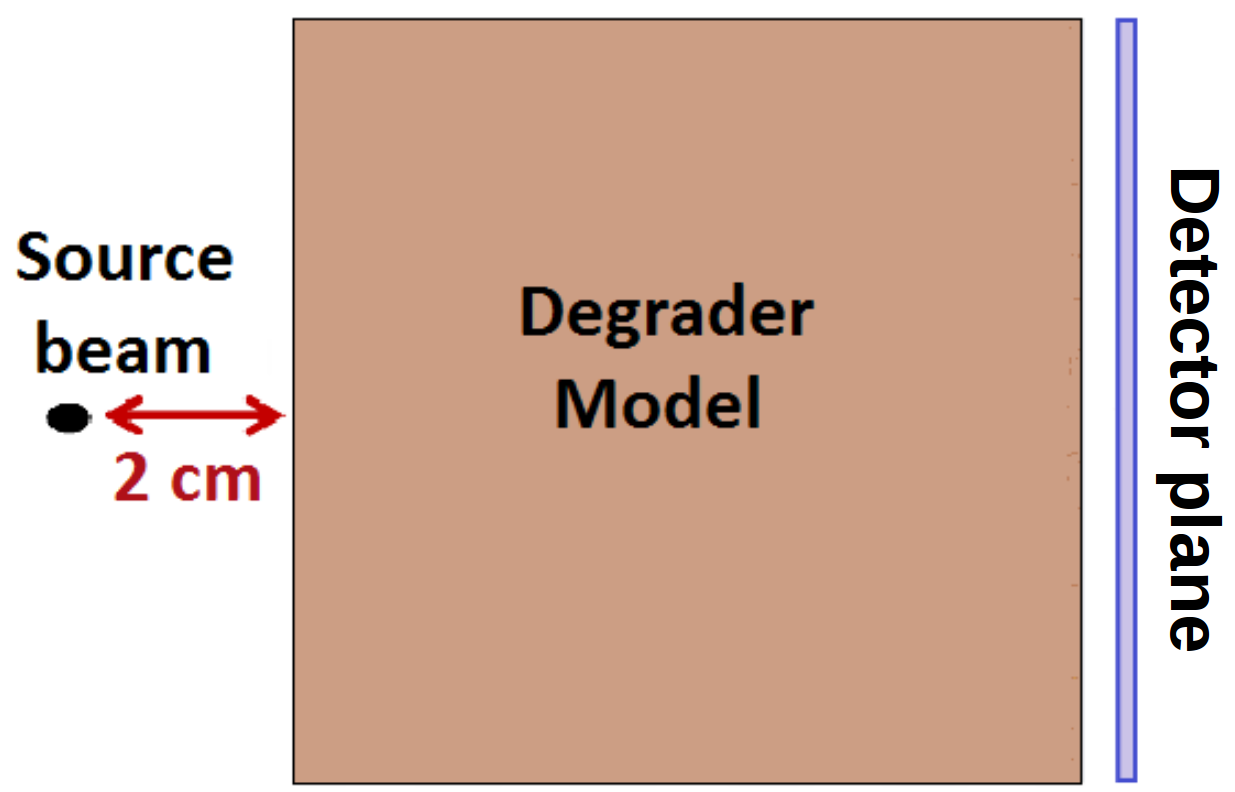}
  \caption{Sketch of the model setup for the degrader simulation.} 
  \label{fig:setup_degr}
\end{figure}

Here the model setup is quite simple:\ an ideal proton beam source (see \secref{subsec:beam}) interacts with the graphite degrader placed 2~cm downstream.\ The model setup does not include any focusing element.\ This avoids additional complications arising from the use of the magnetic elements in the four MC codes.\ The degraded beam phase space is recorded at the detector plane placed 1~mm after the degrader, as shown in \Figref{fig:setup_degr}.\ Keeping the distance fixed between the source beam and the degrader, five different degrader settings, which correspond to five different final energies, are analyzed.

In the following subsections, the main components of the model setup are explained in detail. 

\subsection{Proton beam source}
\label{subsec:beam}

In this work, an ideal proton beam with the parameters of \tabref{table:beampar} is used.\ The choice of an ideal beam with zero transverse divergence and hence zero transverse emittance is motivated by the evaluation of the emittance growth only due to the particle-matter interaction.

\begin{table}[h!]
  \caption{Parameters of the ideal input beam}
     \begin{tabular}{l|r}
      \hline
      \textbf{Parameter} & \textbf{Value} \\	
      \hline
      Number of particles & $10^7$ \\
      Initial kinetic energy     & 249.49 MeV \\
      Transv. distribution type & Gauss\\
      Transv. spatial distribution (FWHM) & 2.35 $\mu$m\\
      Transv. angular distribution (FWHM) & 0 mrad\\
      Energy spread & 0 MeV\\
      Longitudinal bunch length & 0 cm \\
      \hline
    \end{tabular}
  \label{table:beampar}
\end{table}

The initial sample is filled with $10^7$ particles:\ this assures good statistics for the MC simulations in a reasonable computational time. 

\subsection{Degrader}
\label{subsec:degrader}

\begin{figure}[h!]
  \subfloat[Real layout:\ wedges \cite{Reist2002}]{\includegraphics[scale=0.22, keepaspectratio=true]{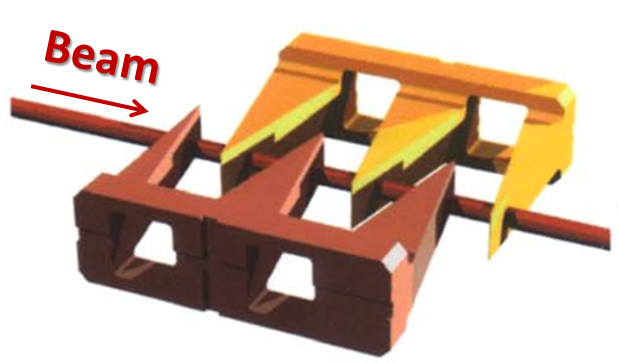}\label{fig:degrader}} \hspace{0.4cm}
  \subfloat[Model geometry:\ slabs]{\includegraphics[scale=0.04, keepaspectratio=true]{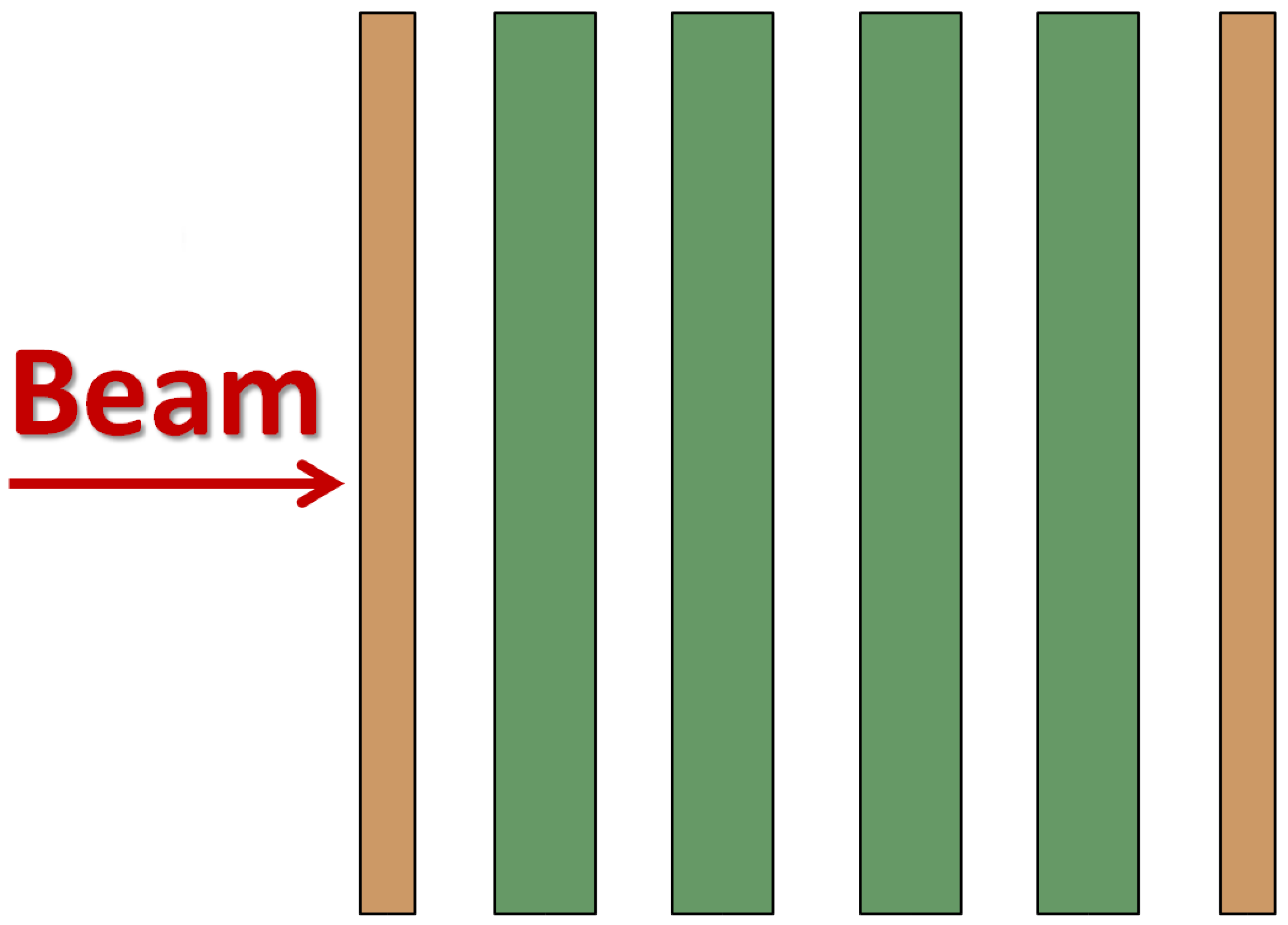}\label{fig:Deg_slab}} 
  \caption{PROSCAN degrader:\ wedge and slab geometry.\ The colors of the slabs underline that the outer slabs (in orange) have half of the thickness of the inner slabs (in green).} 
  \label{fig:PSIdegr}
\end{figure}

As mentioned before, the PROSCAN degrader consists of two pairs of three movable wedges of graphite with a density of 1.88 g/cm$^3$ (\Figref{fig:degrader}).\ Moving the wedges increases or reduces the thickness of graphite that the beam encounters.\ In less than 50~ms any proton beam energy in the range of 230$-$70~MeV is delivered with an accuracy of $\pm 0.1$~mm water-equivalent \cite{Reist2002}.

In the model setup, a simplified geometry of the degrader is implemented with rectangular slabs in place of the wedges (\Figref{fig:Deg_slab}). 

Five different degrader settings that correspond to five final energies between 230 and 70~MeV are simulated.\ For each setting, the wedge position is converted into the equivalent slab thickness.\ The length of the drift space between the slabs is also consequently adjusted.\ As in the real degrader layout (\Figref{fig:degrader}), the first and last slab have half of the thickness of the inner slabs.\ The degrader parameters for the five settings are given in \tabref{table:degrpar}.\

The transverse extension of the slabs (perpendicular to the beam direction) is set to $\pm 40$~cm.\ In this way all scattered particles remain inside the degrader.

\begin{table}[h!]
  \caption{Parameters of the degrader settings.}
     \begin{tabular}{l|c|c|c}
      \hline
      \textbf{Degrader} & \textbf{Inner slab}     & \textbf{Drift}       & \textbf{Tot. graphite} \\	
      \textbf{setting}  & \textbf{thickness}      & \textbf{space}       & \textbf{thickness} \\
      \textbf{[MeV]}    & \textbf{[mm]}           & \textbf{[mm]}        & \textbf{[mm]} \\
      \hline
      230	        & 5.10                   & 36.01               & 25.50 \\
      200               & 13.34                  & 27.77               & 66.72 \\
      150               & 25.45                  & 15.66               & 127.24 \\
      100               & 35.19                  & 5.92                & 175.97 \\
      70                & 39.44                  & 1.67                & 197.19 \\
      \hline
    \end{tabular}
  \label{table:degrpar}
\end{table}

\subsection{Detector plane}
\label{subsec:monitor}

The properties of the beam emerging from the degrader are recorded 1~mm after the last slab (see \Figref{fig:setup_degr}).\ At this position, the transverse phase space and the energy of each particle are collected and used for the analysis.\ The transverse extension of the detector plane is also set to $\pm 40$~cm, as for the degrader slabs.\ This ensures that the scattered 
particles emerging from the degrader are included in the analysis.

\section{The four Monte Carlo codes}
\label{sec:codes}

In the following sections, the four MC codes are briefly described, with particular focus on the features used in the development of the models.

\subsection{FLUKA}
\label{ssec:fluka}

FLUKA is a fully integrated MC simulation code used in a wide range of applications (e.g.\ high energy physics, shielding, cosmic ray studies, medical physics) \cite{Fluka}.\ The FLUKA geometry applies the combinatorial geometry to bodies and region and on their assigned materials.\ For each slab of the degrader model, a corresponding region is defined and the material \textit{graphite} with the specific density of 1.88 g/cm$^3$ assigned.\ A region, in vacuum, is added to the model 1~mm downstream of the last slab and mimics the detector plane.

The proton beam interaction with the degrader is based on the physics model called \textit{PRECISIO}.\ The energy loss is evaluated from the Bethe-Bloch theory with low-energy correction from Ziegler and density effect from Sternheimer \cite{Battistoni2016}.\ The threshold for the particle transport is set at 100 keV.\ The algorithm of the multiple Coulomb elastic scattering is based on Moliere's theory improved by Bethe \cite{Ferrari1992}.\ The inelastic scattering cross-section for hadron-nucleus interactions are obtained from parametrised fits of available experimental data \cite{Fluka}.

In order to record the phase space at the detector plane, the USERDUMP card is used in combination with the FORTRAN routine \textit{mgdraw}, properly adapted to our needs.\

\subsection{GEANT4}
\label{ssec:geant}

GEANT4 is a software toolkit for the simulation of the passage of particles through matter.\ It is used by a large number of experiments and projects in a variety of application domains, 
including high energy physics, astrophysics and space science, medical physics and radiation protection~\cite{Geant4}.\ 

Since GEANT4 is a library and not a standalone program, an additional program needs to be used. For this work the open source C++ BDSIM tracking code is chosen (version 0.992 with GEANT4 version 10.2.2)~\cite{BDSIM}.

The degrader model of \secref{sec:model} is modeled in BDSIM as follows.\ For the slabs rectangular collimators with a zero aperture are used.\ For the detector plane BDSIM has the so-called \textit{sampler} element that records all particles that pass it.

The \textit{G4EmStandardPhysics} physics model was activated in this work.\ It includes the energy loss based on the Bethe-Bloch equation with Barkas and Sternheimer corrections and the multiple elastic scattering based on Lewis theory \cite{Lewis1950} instead of the Moliere formalism.\ To include also the inelastic scattering in the physics model, the \textit{G4HadronPhysicsQGSP\_BIC\_HP} process was activated.\ To cut on low energy particles, in GEANT4 one has to ``cut in range''.\ The proton production cut is set to 1~mm, which corresponds to a proton energy cut of about 100~keV.

\subsection{MCNPX}
\label{ssec:mcmpx}

MCNPX is a general purpose MC code that allows for simulating the transport through matter of more than 30 types of particles and ions over an energy range from $\sim$\,10$^{3}$ GeV per nucleon down to 10$^{-11}$~MeV \cite{MCNPX}.\ The MCNPX version 2.7.0 with the default set of physics settings was used in this study.\ In particular MCNPX utilizes the condensed history algorithm, computing stopping powers for the charged particles using the Bethe-Bloch formula with Sternheimer and Peierls density effect correction and accounting for multiple elastic scattering with the modified Rossi theory and energy straggling with Vavilov's model for heavy charged particles.\ To describe the proton-nucleus interactions, MCNPX uses the intranuclear cascade model of Bertini coupled with the Dresner evaporation model.

The MCNPX geometry consists of three-dimensional cells defined through bounding the cells by the surfaces of the first and second order.\ Thus a seamless realization of the geometrical model based on a similar approach, like the FLUKA geometry model, is straightforward.\ The MCNPX capabilities allow for automatized dumping of the state of particle transport events basing either on the cell or on the surface selection and filtering.\ For the presented MCNPX simulations, the model setup described in \secref{sec:model} has been implemented and the parameters of particle tracks are recorded at the detector plane transverse to the beam direction.

\subsection{OPAL}
\label{ssec:opal}

OPAL is an open-source three-dimensional tracker for general particle accelerator simulations \cite{opal:1}.\ It has the unique feature to combine seamless linear and nonlinear beam tracking with MC simulations of the particle-matter interaction. 

When a particle hits one of the graphite slabs, OPAL evaluates the energy loss and elastic scattering angle \cite{Stachel2013} using more simplified models with respect to the fully integrated MC codes.\ In particular, for the proton energy loss, the empirical formula from Andersen and Ziegler \cite{Andersen} is used for energies below 0.6~MeV and the Bethe-Bloch equation \cite{PDG} for energy above it.\ In terms of the elastic scattering, both multiple Coulomb scattering and single Rutherford scattering are available following the implementation in \cite{PDG,Jackson}.\ Finally, the particle-matter interaction in the current version of OPAL is restricted to protons and inelastic nuclear interactions are not implemented.\ In contrast to the fully integrated MC codes, the slabs and the detector plane in OPAL have an infinite transverse extension.\ However, this fact does not influence the correctness of the comparison with the other MC codes since the chosen finite extension ($\pm 40$~cm) is large enough to enclose the entire beam.

While in the conventional MC codes the step size for particle transport is internally optimised, in OPAL it has to be set by the user in the input file.\ In this work, a step size of 0.2~mm is used such that particle 
velocity remains essentially constant during the step.

\section{Validation of the slab geometry}
\label{sec:geo}

The use of multi slabs instead of wedges (\Figref{fig:PSIdegr}) is motivated by the OPAL limitation in the degrader geometry.\ Before developing the degrader models, we verified the correctness of the slab geometry approximation.\ In particular, we analysed the 230~MeV degrader setting, where the wedge shape is expected to have a bigger influence. 

\begin{figure}[h!]
  \centering
  \includegraphics[scale=0.13, keepaspectratio=true]{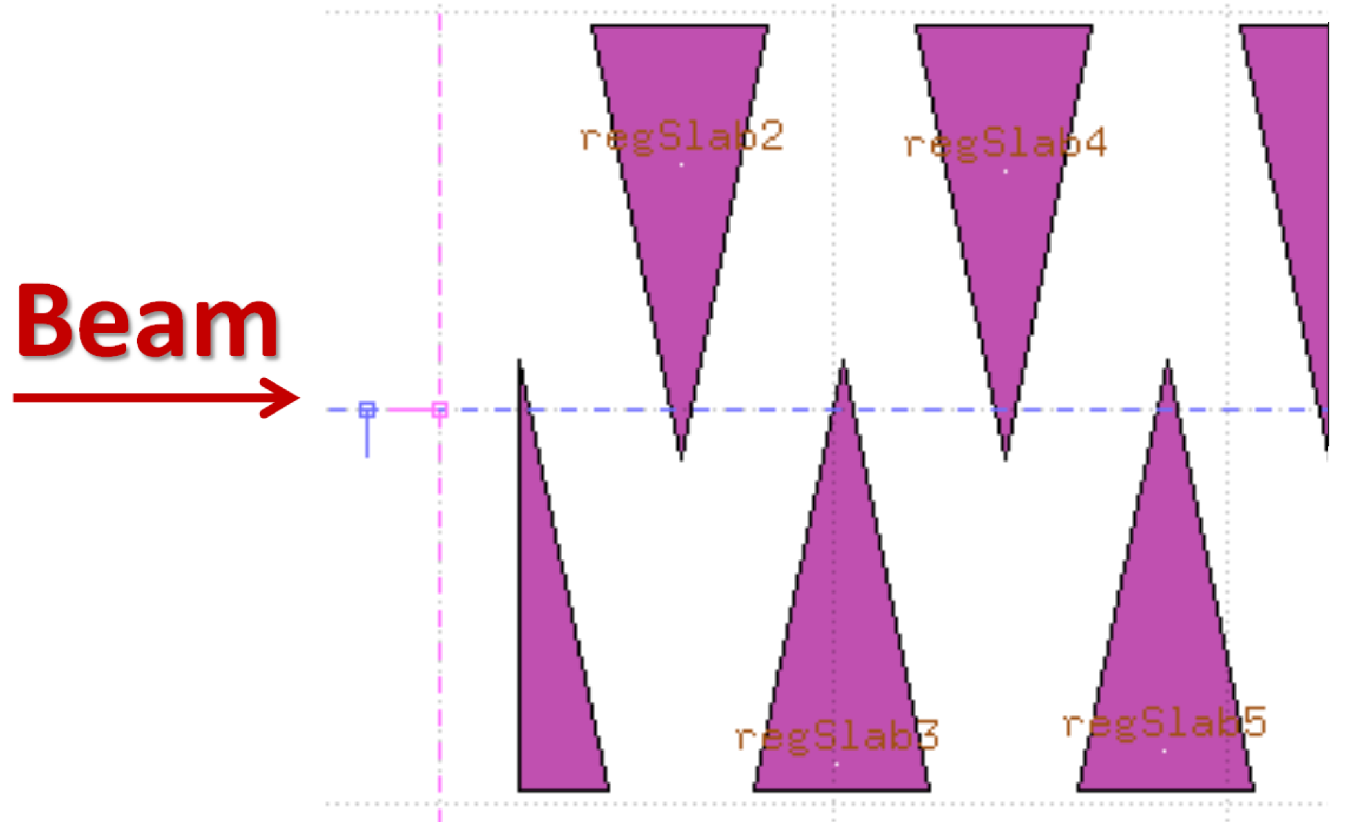}
  \caption{Top-view of the wedge geometry implemented in FLUKA for the 230~MeV degrader setting.}
  \label{fig:Wedge_geo}
\end{figure}

FLUKA allows high flexibility in the definition of the degrader geometry.\ For this reason, the wedge geometry of the degrader as in \Figref{fig:degrader} has been implemented.\ Knowing the vertex angle (23$^\circ$ for the inner wedges) and the position of the wedges with respect to the incoming beam, the 230~MeV degrader setting geometry is simulated in FLUKA as shown in \Figref{fig:Wedge_geo}.\  
 
\begin{figure}[h!]
  \centering
  \includegraphics[scale=0.22, keepaspectratio=true]{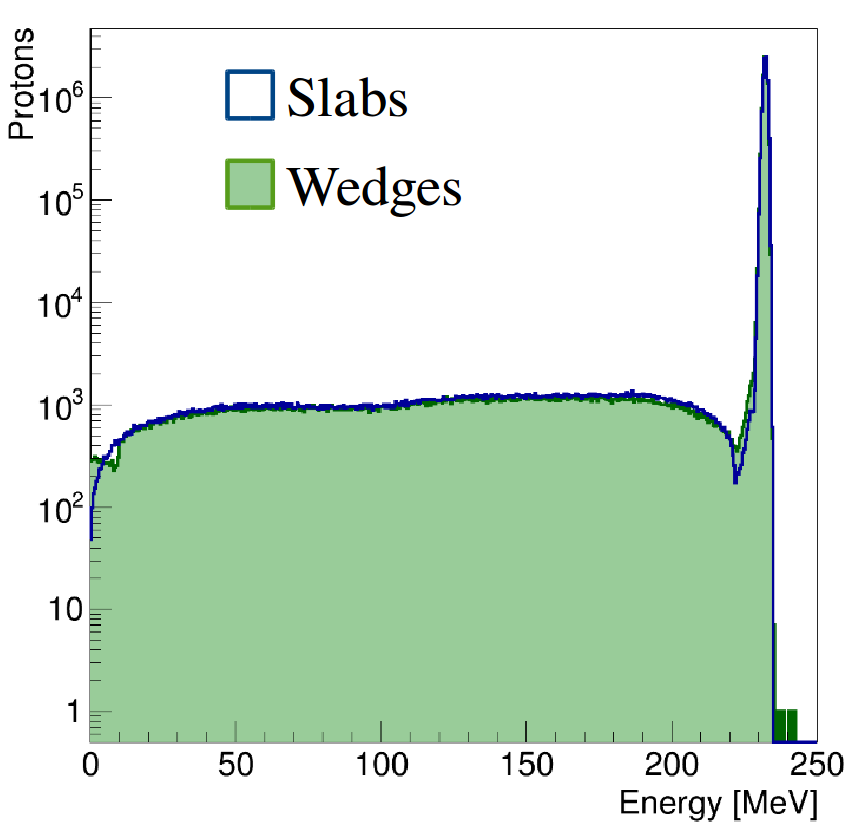}
  \caption{Energy spectrum comparison between the multi slabs and the wedge geometry.} 
  \label{fig:Spectrum_Wedge}
\end{figure}

The detector plane placed 1~mm after the last wedge records the degraded proton beam.\ The comparison between the two geometries (slab and wedge) is performed in terms of the energy spectrum (\Figref{fig:Spectrum_Wedge}).

The perfect agreement between the two energy spectra validates the simplification adopted in the slab geometry.

\section{Degrader models: results}
\label{sec:results}

\begin{figure}[h!]
 \centering
 \includegraphics[scale=0.16, keepaspectratio=true]{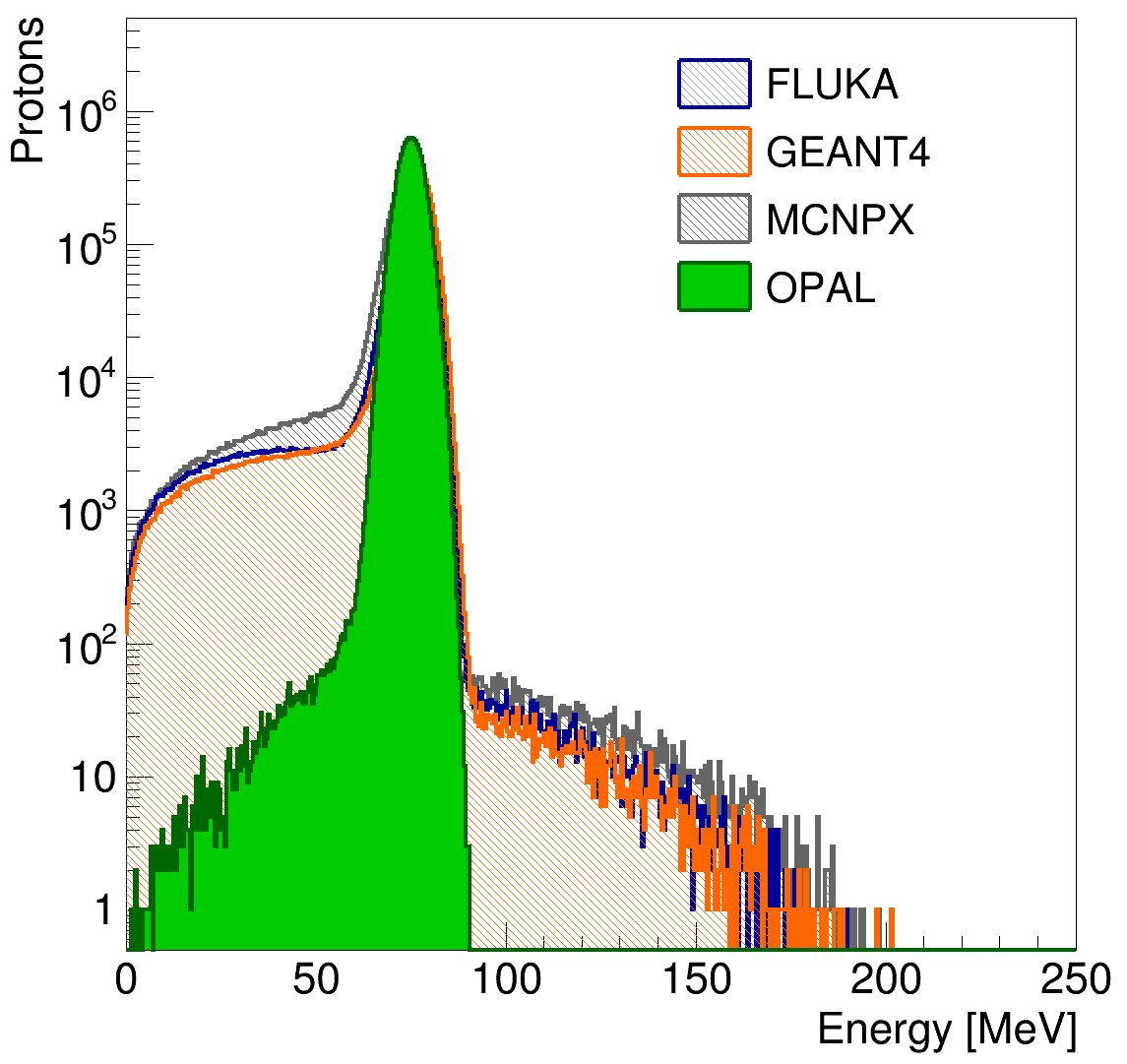}
  \caption{Energy spectra from the four MC codes for the 70~MeV degrader setting.\ The main peak is surrounded by tails due to inelastic scattering, except for OPAL where this process is not implemented.} 
  \label{fig:Spectrum70}
\end{figure}

The interaction of the proton beam with the degrader is dominated by the multiple Coulomb scattering (elastic and inelastic) that spreads the phase space of the beam and creates energy tails.\ A typical degraded energy spectrum shows a main peak followed by inelastic tails.\ \Figref{fig:Spectrum70} displays a comparison between the energy spectra from the four MC codes for the 70~MeV degrader setting.

The agreement between the FLUKA, GEANT4 and MCNPX spectra is qualitatively good, with only a small discrepancy in the inelastic tail.\ The comparison with the OPAL spectrum can be performed only on the main peak, since (as already mentioned) the inelastic scattering is not implemented.

In the following subsections, the main results from the comparison of the four codes are reported.\ The analysis is focused on the parameters that are important for the development of the transport line model downstream of the degrader. 


\subsection{Transmission through the degrader}
\label{ssec:trasmission}

\begin{figure}[h!]
 \centering
 \includegraphics[scale=0.16, keepaspectratio=true]{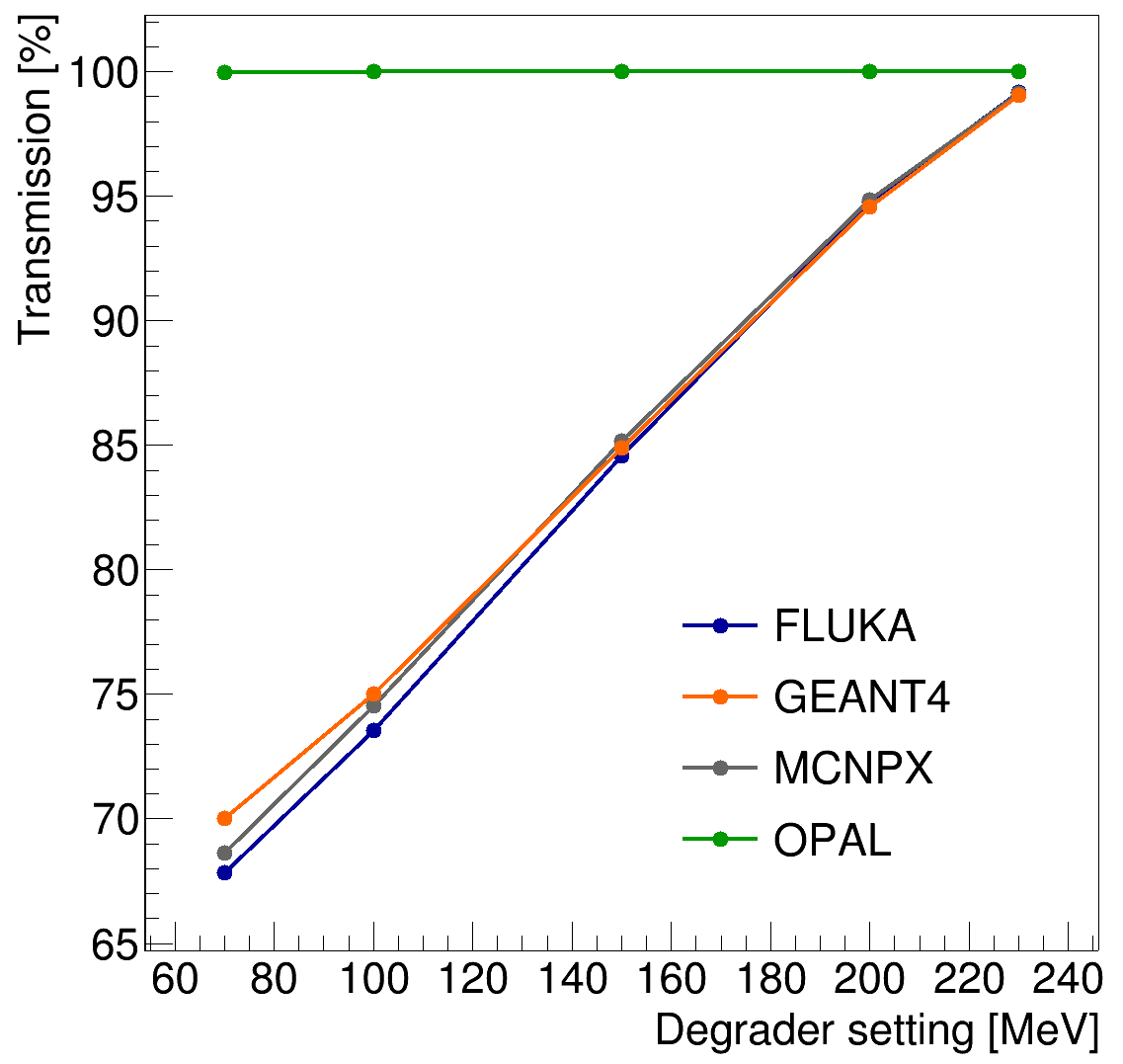}
  \caption{Transmission through the degrader:\ comparison of the four MC codes.} 
  \label{fig:transmission}
\end{figure}

In this analysis, we are interested in the fraction of the initial protons that pass through the degrader and reach the detector plane.\ This estimates the losses in the degrader due to proton absorption or very large angle scattering outside the detector plane.\ We define the transmission through the degrader as

\begin{equation}
  \rm{Transmission} \hspace{0.1cm} [\%] = {N_\text{tot}}/{N_\text{ini}} \times 100
\end{equation} 
  
where N$_\text{ini}$ is the initial number of protons in the source beam ($10^7$ from Table \ref{table:beampar}) and N$_\text{tot}$ indicates the total number of protons recorded at the detector plane.\ The transmission from the four MC codes for the five degrader settings is shown in \Figref{fig:transmission}. 

The results of the fully integrated MC codes (FLUKA, GEANT4 and MCNPX) are in a good agreement and show a reduction of up to 30\% in the transmission when lowering the beam energy, i.e.\ the increasing of the degrader thickness.\ Since OPAL does not account for inelastic scattering, almost full transmission (close to 100\%) is obtained.\ The remaining tiny losses in OPAL are related to the OPAL-internal energy threshold for particle transport, set to 100~keV.\ A possible compensation for the missing of a inelastic scattering module in OPAL is discussed in \secref{ssec:IS_Appl}.

To verify the correctness of the OPAL results of \Figref{fig:transmission}, the inelastic scattering effect has been disabled in FLUKA and the simulation of the five degrader settings repeated.\ The obtained transmission values are reported in \tabref{table:transmission}.  
      
\begin{table}[ht!]
  \centering
  \caption{Transmission [\%] through the degrader:\ comparison between OPAL and FLUKA with enabled/disabled the inelastic scattering.}
     \begin{tabular}{l|c|c|r}
      \hline
      \textbf{Degrader}   & \multicolumn{2}{c|}{\textbf{FLUKA}} & \multirow{2}{*}{\textbf{OPAL}} \\
      \hhline{~--~}
      \textbf{Setting}   & \textbf{Enabled}  & \textbf{Disabled} & \\
       \hline
	230 MeV  	& 99.16 & 99.99 & 99.99 \\
	200 MeV  	& 94.79 & 99.99 & 99.99 \\
	150 MeV  	& 84.58 & 99.99 & 99.99 \\
	100 MeV  	& 73.54 & 99.97 & 99.99 \\
	 70 MeV  	& 67.85 & 99.89 & 99.98 \\
        \hline
    \end{tabular}
  \label{table:transmission}
\end{table}

When the inelastic scattering effect is disabled in FLUKA, the discrepancy with the OPAL results is less than 0.1\%.\ 

\subsection{Beam energy and energy spread}
\label{ssec:energy}

\begin{figure}[h!]
 \centering
 \includegraphics[scale=0.14, keepaspectratio=true]{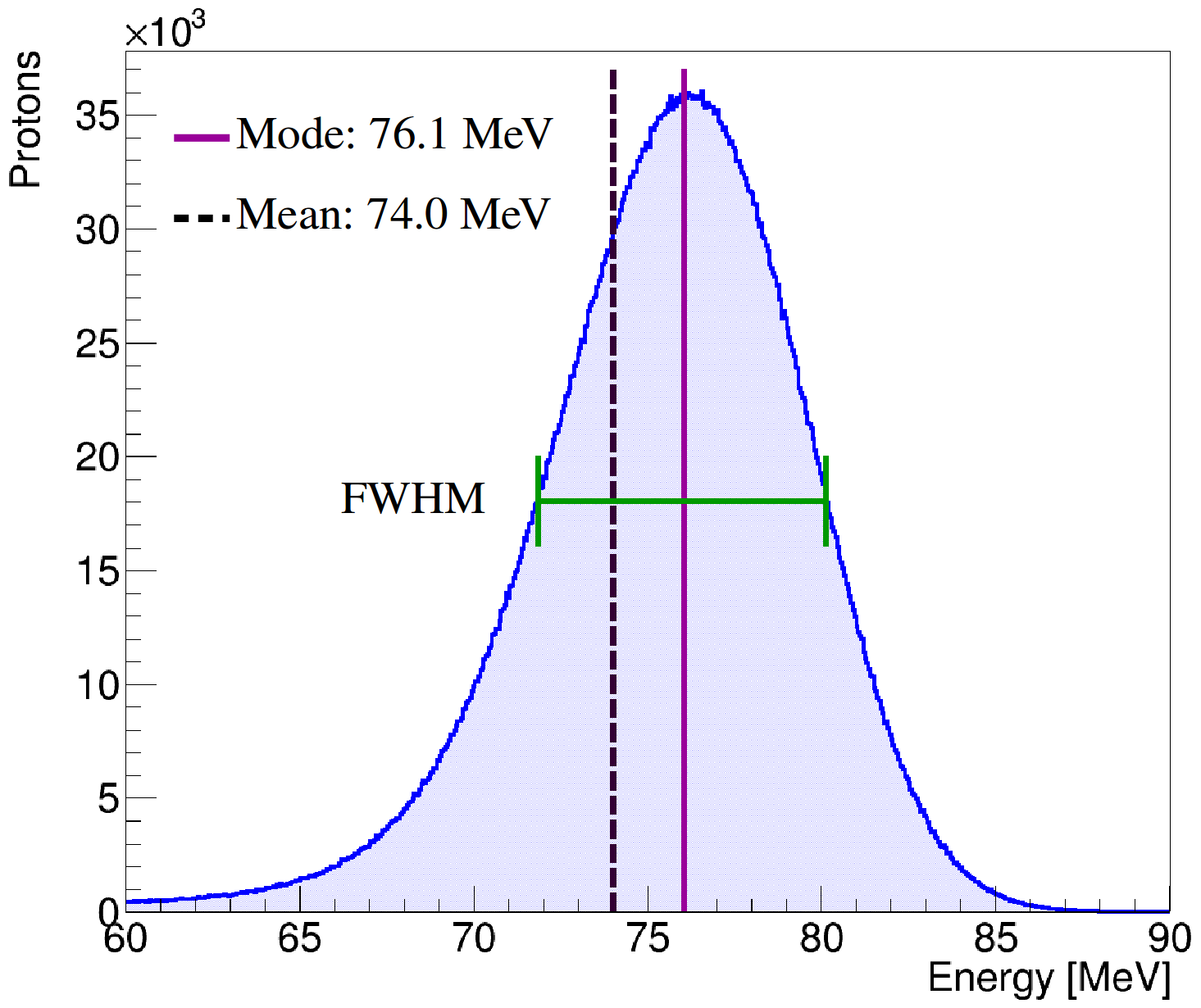}
 \caption{Beam energy analysis:\ mean or mode from statistics.\ The energy spread is obtained from the FWHM (green line) with respect to the mode of the energy distribution.}
 \label{fig:SpectrumAnalysis}
\end{figure}

Beam energy and energy spread determine the penetration depth and dose distribution within the tumor tissue and have to be controlled with high accuracy.\ There are several methods to calculate the beam energy from the degraded energy spectra (\Figref{fig:Spectrum70}).\ The first approach is to analyse the main peak with a Gaussian fit and use the mean fit value as beam energy.\ However, the accuracy of the Gaussian fit is reduced, especially at the low energy degrader settings, due to the skewness of the distribution.\ In the second approach the statistical properties (mean and mode) are used.\ Since the mean value (dashed black line in \Figref{fig:SpectrumAnalysis}) is influenced by the presence of the low energy inelastic tails, in our analysis we use the mode (i.e.\ the energy interval containing the maximum number of protons) of the energy distribution to define the beam energy $\bar{E}$ (purple line in \Figref{fig:SpectrumAnalysis}).\ The Full Width Half Maximum (FWHM) of the main peak with respect to $\bar{E}$ is used to evaluate the energy spread.\ In particular the energy spread is defined as the standard deviation $\sigma_E$ associated to the FWHM ($\sigma_E$ = FWHM/2.355).

The beam energy from the mode is calculated for the five degrader settings and the resulting values from FLUKA, used as reference, are reported in \tabref{table:FlukaEne}.\

\begin{table}[ht!]
 \centering
  \caption{Beam energy $\bar{E}$ and energy spread $\sigma_E$ from FLUKA for the five degrader settings.}
     \begin{tabular}{l|c|c}
      \hline
      \textbf{Degrader} & \textbf{Beam energy} & \textbf{Energy spread} \\
      \textbf{Setting} &  \textbf{[MeV]} &  \textbf{[MeV]}  \\
       \hline
	230 MeV  	& 232.51 & 0.70  \\
	200 MeV  	& 202.94 & 1.19 \\
	150 MeV  	& 153.74 & 1.85 \\
	100 MeV  	& 104.16 & 2.71 \\
	 70 MeV  	& 76.06  & 3.49 \\
        \hline
    \end{tabular}
  \label{table:FlukaEne}
\end{table}

\begin{figure}[ht!]
  \centering
  \subfloat[Beam energy $\bar{E}$]{\includegraphics[scale=0.16, keepaspectratio=true]{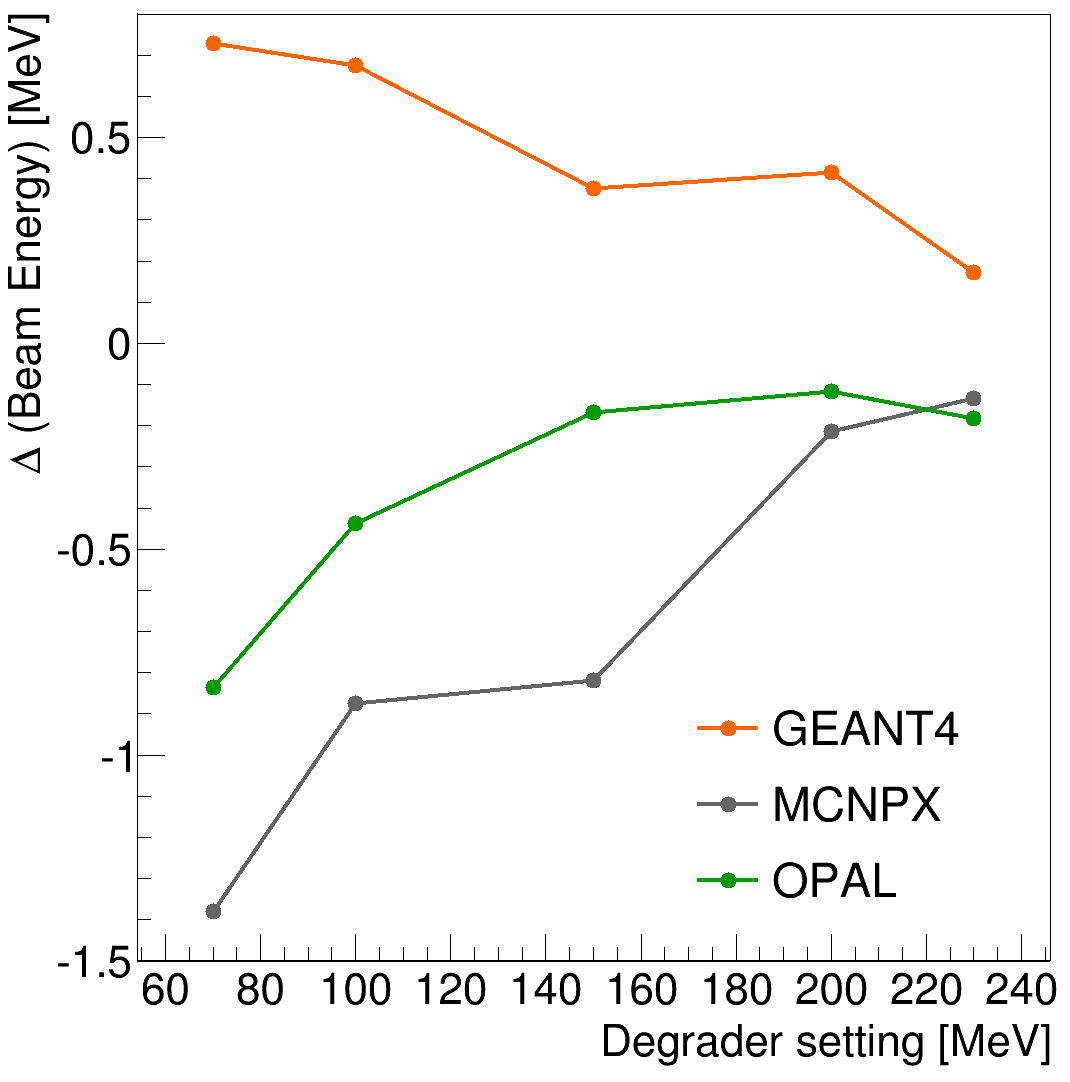}\label{fig:EnergyComp}}\\
  \subfloat[Energy spread $\sigma_E$]{\includegraphics[scale=0.16, keepaspectratio=true]{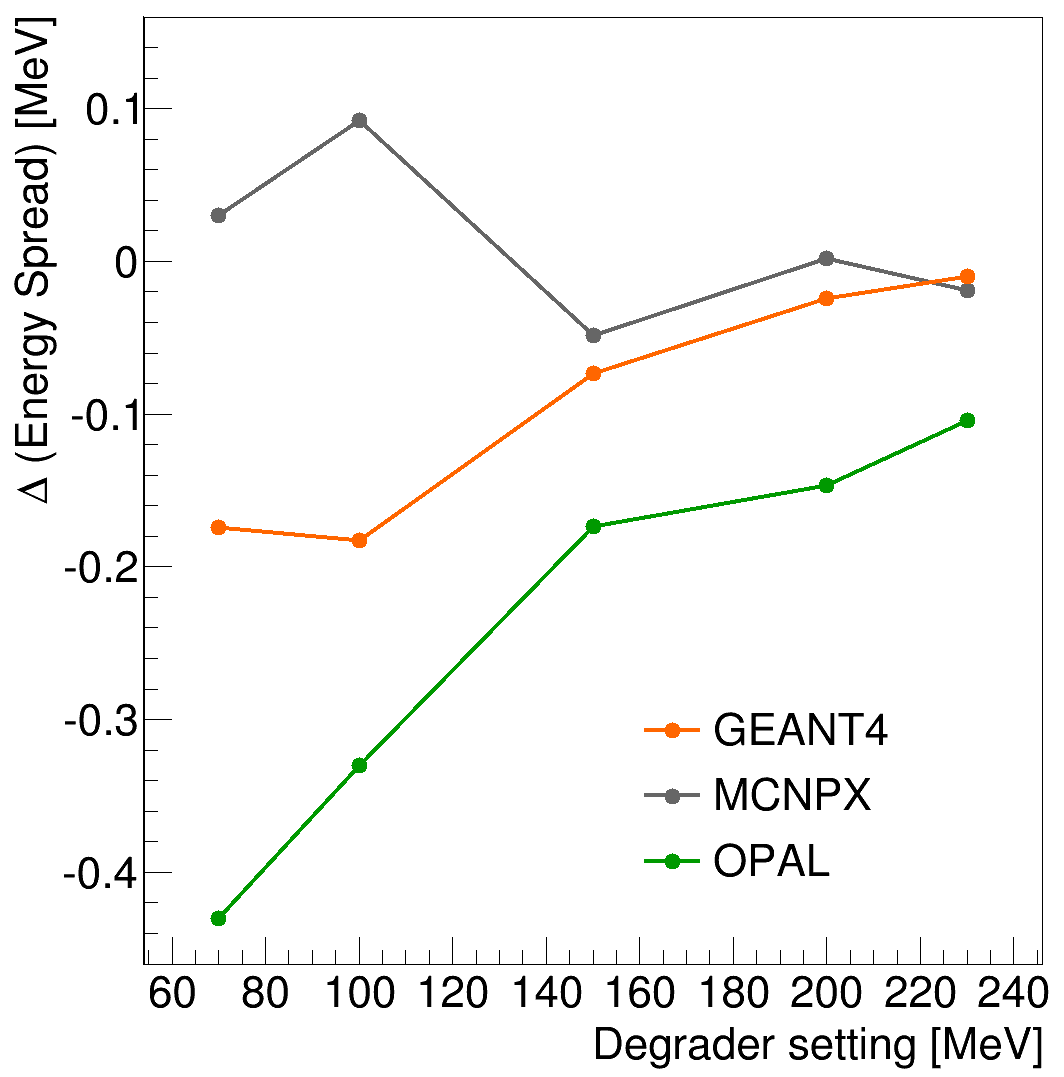}\label{fig:EnergySpreadComp}} 
  \caption{Difference between beam energy and energy spread from three MC codes (GEANT4, MCNPX and OPAL) and from the FLUKA data of \tabref{table:FlukaEne} considered as reference.} 
  \label{fig:MeanAndSpread}
\end{figure}

The same procedure is repeated for the other MC codes (GEANT4, MCNPX and OPAL) and their results compared with the FLUKA data of \tabref{table:FlukaEne}, as displayed in \Figref{fig:MeanAndSpread}.

For the degrader setting at higher energies, the discrepancy in the beam energy between the four MC codes is below $\pm$0.1\%.\ The good agreement of OPAL with the general MC codes validates the correct implementation of the simplified Bethe-Bloch equation that neglects the density correction \cite{opal:1}.\ The maximum discrepancy in the beam energy is 1.8\% between FLUKA ($\bar{E}$ = 76.06~MeV) and MCNPX ($\bar{E}$ = 74.68~MeV).\ At 70~MeV degrader setting the main peak is quite broad, left-skewed and with the top part flat.\ The mode of the energy distribution can be shifted from the ideal central position resulting in a bigger discrepancy with respect to the other codes.\ Several bin widths have been tested to verify how the beam energy is influenced and a final bin width of 0.005~MeV has been chosen without loss in accuracy.

For the energy spread, the bigger discrepancies (12\%~between OPAL and FLUKA) appear also at 70~MeV degrader setting.\ However, this discrepancy has to be compared with the small aperture of the horizontal slit in the ESS.\ Especially at low energies, in fact, the energy spread of the beam is defined by the ESS.\ Additional details are given in \secref{ssec:Ene_AP} where we discuss the computed results for the energy spread in relation to the ESS settings normally used in a proton therapy facility.

\subsection{Beam core and tail: inelastic scattering contribution}
\label{ssec:ineScat}

Cyclotron-based proton therapy facilities are typically equipped with collimators and ESS right after the degrader to limit the transmitted beam to the acceptance of the beam transport system.\ Here also the inelastic tail is removed and only the co-called beam core is left.\ 

In order to evaluate the contribution of the inelastic scattering, we distinguish the beam core from the tails.\ The three fully integrated MC codes utilize quite different intrinsic definitions for what shall be regarded as beam core.\ In order to regularise the analysis, a standardised method to define the beam core has been developed and applied, in the same way, to the degraded energy spectra obtained from FLUKA, GEANT4 and MCNPX. 

In particular, the beam energy $\bar{E}$ and the energy spread $\sigma_E$, as defined in \secref{ssec:energy}, are used to identify the ensemble of protons that belong to the beam core ($\varepsilon_\text{core}$) and the ensemble of protons in the tails ($\varepsilon_\text{tail}$).\ The total ensemble of protons $\varepsilon_\text{tot}$ that reach the detector plane is formed by:\ $\varepsilon_\text{tot} = \varepsilon_\text{core} \cup \varepsilon_\text{tail}$.\ In terms of the number of protons that form these ensembles, we have:\ N$_\text{tot}$ = N$_\text{core}$ + N$_\text{tail}$.\ In particular, a proton with energy $E$ belongs to the ensemble $\varepsilon_\text{core}$ if

\begin{equation}
  \bar{E} - 3 \sigma_E \leq E \leq \bar{E} + 3 \sigma_E.
  \label{eq:range}
\end{equation} 

\begin{figure}[h!]
  \centering
  \includegraphics[scale=0.10, keepaspectratio=true]{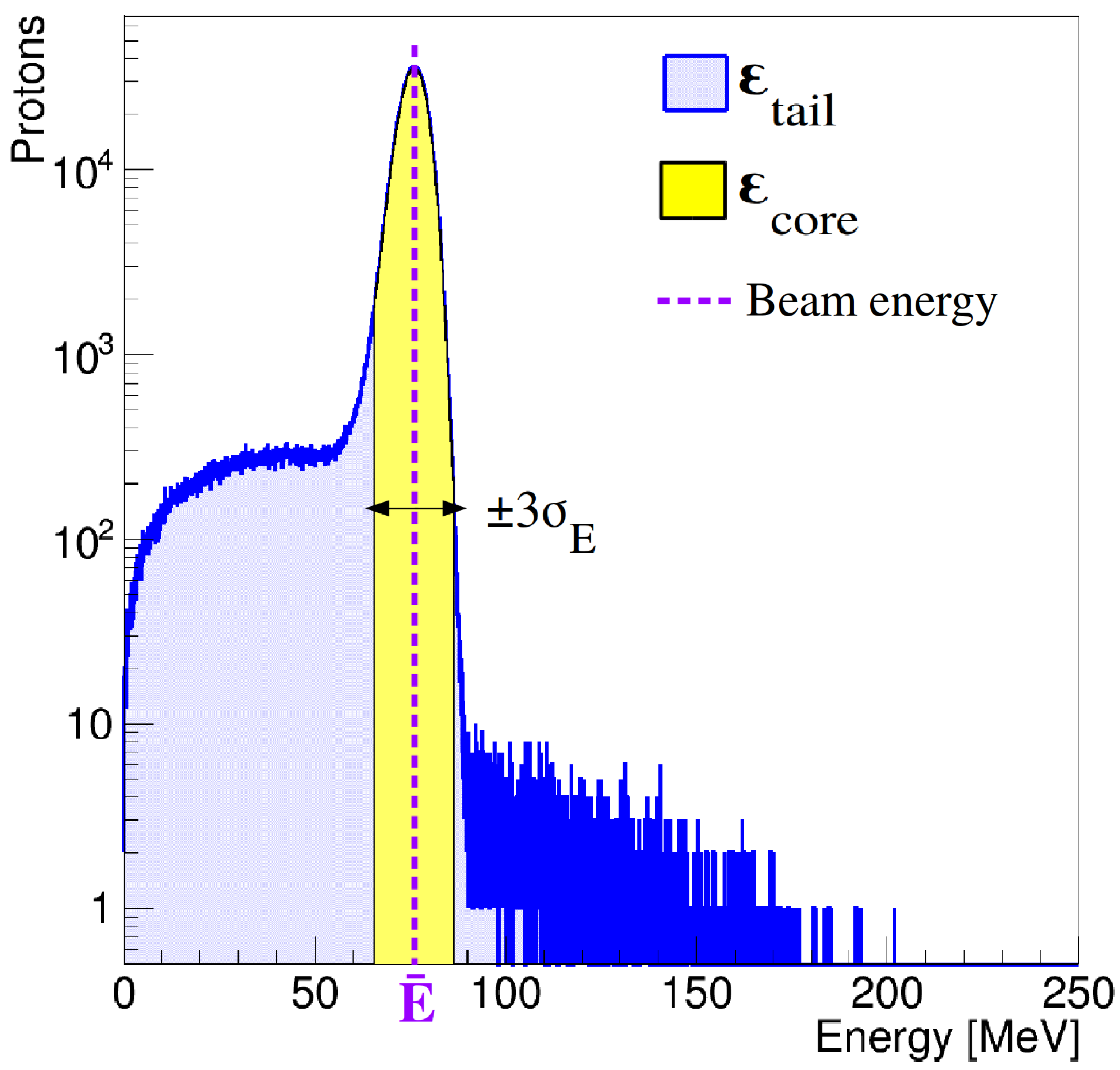}
  \caption{Analysis of a degraded energy spectrum: the beam energy $\bar{E}$ and energy spread $\sigma_E $ are used to distinguish the core of the beam ($\varepsilon_\text{core}$) inside $\bar{E}\pm 3\sigma_E $ from the tail ($\varepsilon_\text{tail}$) outside this range.} 
  \label{fig:SpectrumFit}
\end{figure}

The use of $\pm 3\sigma_E$ ensures to populate the ensemble $\varepsilon_\text{core}$ with about 99\% of the particles in the main peak.\ An example of this analysis is shown in \Figref{fig:SpectrumFit}.

Protons outside the energy range of \Eqref{eq:range} belong to the ensemble $\varepsilon_\text{tail}$ and are used to evaluate the contribution of the inelastic scattering.\ In particular, we are interested in the amount of protons (N$_\text{tail}$) that form the ensemble $\varepsilon_\text{tail}$.

The contribution of the inelastic scattering tail over the total number of protons N$_\text{tot}$ is evaluated as 

\begin{equation}
  \rm{Inelastic \hspace{0.1cm} tail} \hspace{0.1cm} [\%] = {N_\text{tail}}/{N_\text{tot}} \times 100.
\end{equation} 

\begin{figure}[h!]
 \centering
 \includegraphics[scale=0.17, keepaspectratio=true]{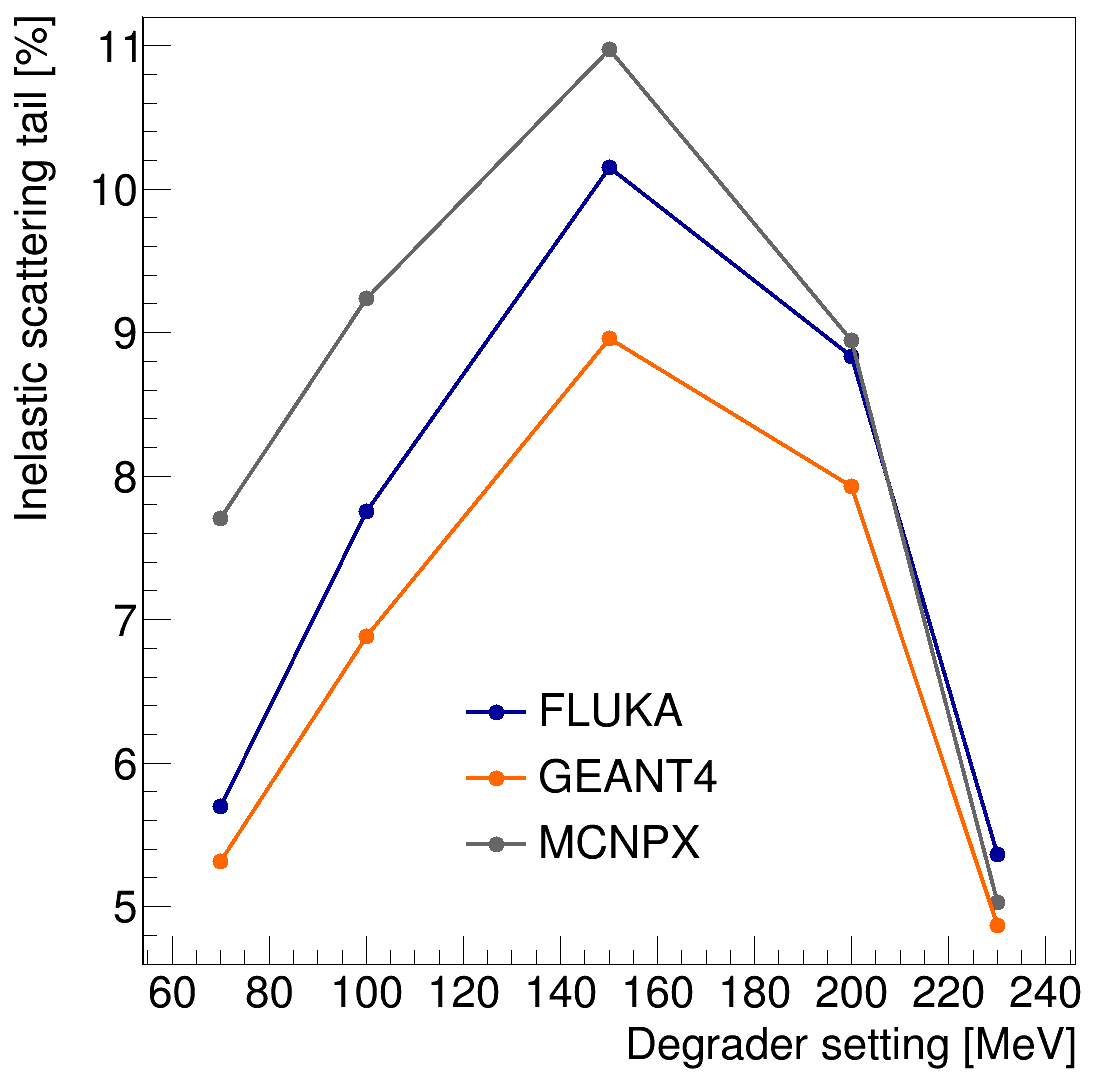}
 \caption{Contribution of the inelastic scattering tail to the full energy spectrum:\ comparison between FLUKA, MCNPX and GEANT4.} 
  \label{fig:inescat}
\end{figure}

This analysis is restricted to the fully integrated MC codes and the results are shown in \Figref{fig:inescat}.\ 

From all codes, the maximum inelastic scattering contribution is reached at 150~MeV degrader setting.\ Lowering the final beam energy and hence increasing the graphite thickness, the inelastic scattering contribution decreases again to 5$-$7\%.\ In the higher energy range (230$-$150~MeV), the inelastic scattering dominates over the proton absorption.\ At 150~MeV, in fact, 85\% of the incoming beam still reaches the detector plane, as visible in \Figref{fig:transmission}.\ For energies below 150~MeV, the proton absorption and the energy loss become dominant effects reducing the contribution from the inelastic scattering.\ This behavior explains the trend of the inelastic scattering contribution shown in \Figref{fig:inescat}.

\subsection{Transverse phase space and emittance growth}
\label{ssec:phasespace}

The proton interaction with the degrader leads to a growth of the transverse phase space and beam emittance.

\begin{figure}[h!]
  \centering
  \includegraphics[scale=0.17, keepaspectratio=true]{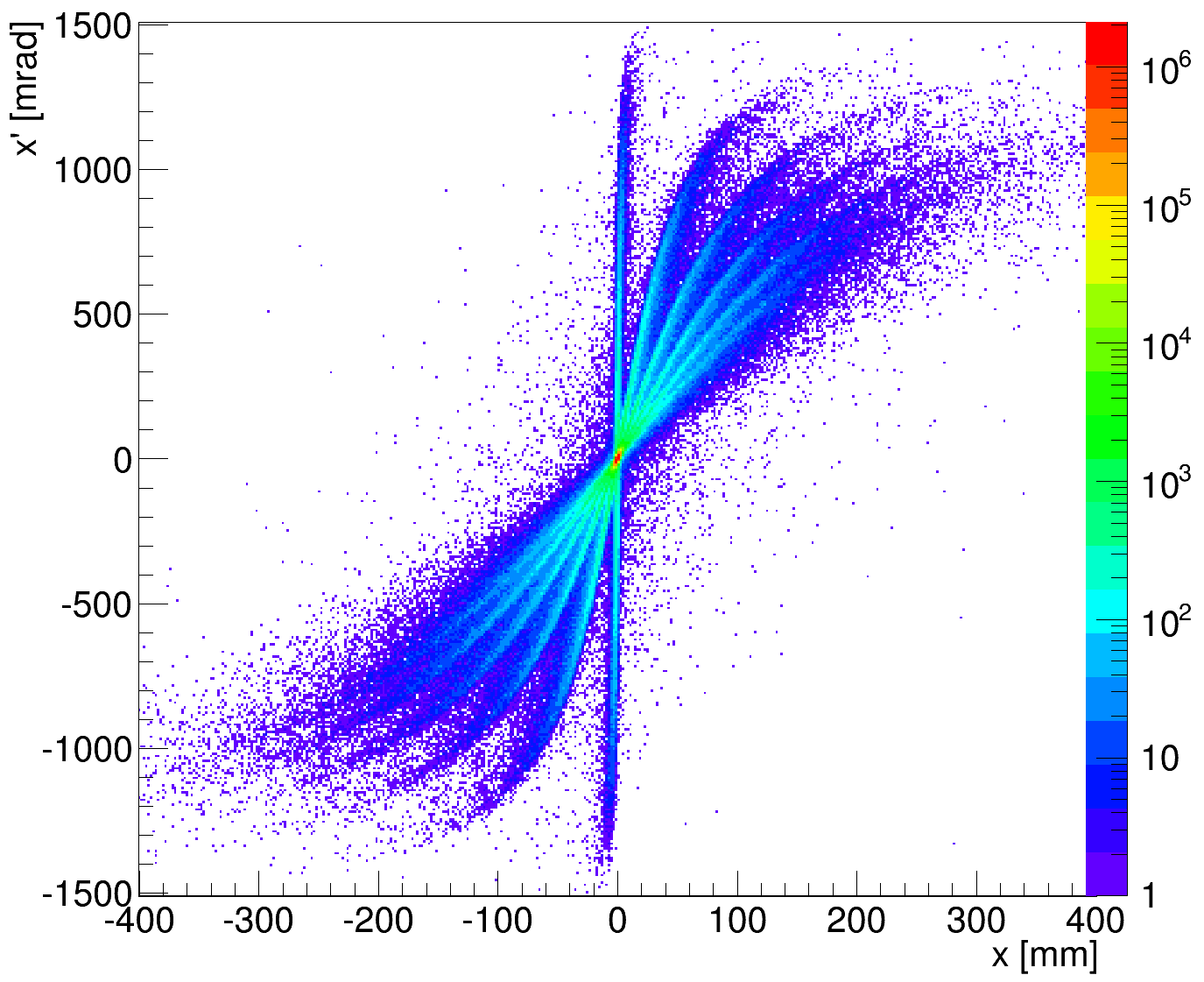}
  \caption{Horizontal phase space of the degraded beam from FLUKA at the detector plane for the 230~MeV degrader setting.} 
  \label{fig:PhaseSpace}
\end{figure}

The use of an ideal source beam with zero transverse emittance allows evaluating the growth of divergence and emittance only due to the particle-matter interaction.\ An example of the transverse (horizontal) beam phase space after the degrader is shown in \Figref{fig:PhaseSpace}. 

In \Figref{fig:PhaseSpace} six structures are visible in the inelastic tail.\ They are particularly evident at the 230~MeV degrader setting, where the graphite slabs are thin (5~mm) and separated by 36~mm of vacuum space.\ The structures simply reflect the effect of the proton scattering within the slabs.\ In fact, after leaving a slab with a significant scattering angle, protons continue their trajectories in the free space before reaching the subsequent slab.\ The shape of central structure (for x = 0~mm) is due to the small distance (1~mm) between the last slab and the detector plane.\ This short space does not allow the scattered beam to expand properly.\ Therefore, the angular projection of the proton trajectories at the detector plane creates this central structure that behaves like the other structures as soon as the detector plane is moved away from the degrader.

\begin{figure}[ht!]
  \centering
  \includegraphics[scale=0.11, keepaspectratio=true]{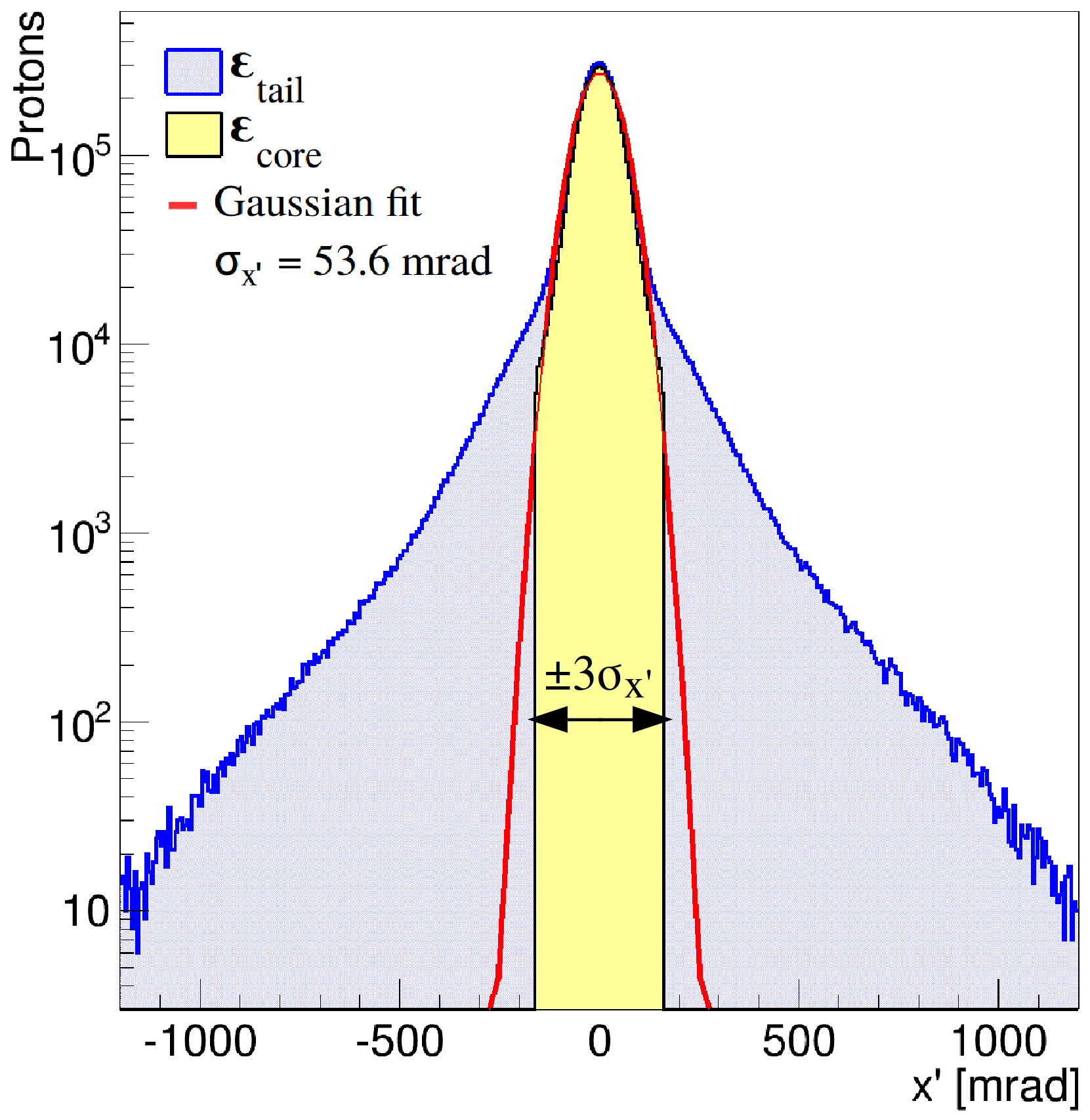}
  \caption{Analysis of the horizontal divergence distribution to distinguish the core of the beam ($\varepsilon_\text{core}$) inside $\pm 3\sigma_{x'} $ from the tail ($\varepsilon_\text{tail}$) outside this range.} 
  \label{fig:PhaseSpaceAnalysis}
\end{figure}

To evaluate the growth of divergence and emittance due to the degradation process, we slightly modify the method described in \secref{ssec:ineScat}.\ As mentioned before, after collimators and ESS, only the beam core is transported along the transport line downstream of the degrader.\ For this reason, we limit the analysis of the increased transverse phase space only to the beam core.\ In this case, the distinction between $\varepsilon_\text{core}$ and $\varepsilon_\text{tail}$ is based on the properties of the transverse phase space (spatial coordinates and divergences distribution) rather than on the energy spectrum, as done in \Figref{fig:SpectrumFit}.\ An example of this analysis is shown in \Figref{fig:PhaseSpaceAnalysis} where the divergence distribution at the detector plane is analyzed with a Gaussian fit.

The width of the Gaussian fit, $\sigma_{x'}$ in case of \Figref{fig:PhaseSpaceAnalysis}, is used to define the ensemble $\varepsilon_\text{core}$ for the horizontal plane together with the corresponding $\sigma_{x}$, width of the Gaussian fit on the spatial-x coordinate distribution.\ In this analysis a proton with coordinates ($x$, $y$) and divergences ($x'$, $y'$) at the detector plane belongs to the $\varepsilon_\text{core}$ if:

\begin{equation}
  \begin{aligned}
 - 3 \sigma_x \leq x \leq + 3 \sigma_x  \quad & \land \quad  - 3 \sigma_y \leq y \leq + 3 \sigma_y \quad \land \\
 - 3 \sigma_{x'} \leq x' \leq + 3 \sigma_{x'}  \quad & \land \quad  - 3 \sigma_{y'} \leq y' \leq + 3 \sigma_{y'}.
 \end{aligned}
 \label{eq:range2}
\end{equation}

\begin{figure}[!ht]
  \centering
  \includegraphics[scale=0.17, keepaspectratio=true]{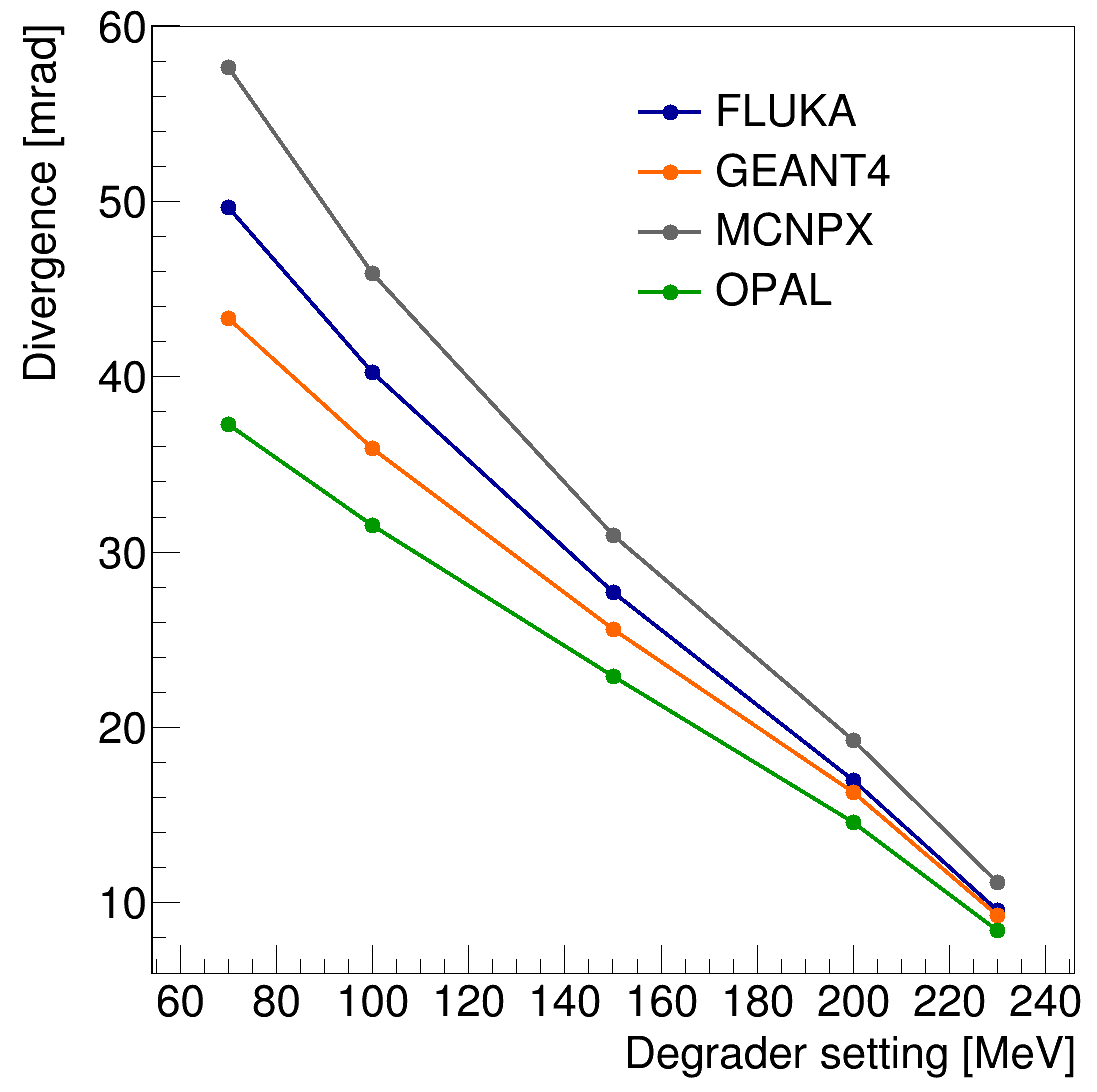}
  \caption{Transverse divergence of the protons in the $\varepsilon_\text{core}$: comparison between the four MC codes.} 
  \label{fig:phasespace}
\end{figure}

The coordinates and divergences of the particles that satisfy \Eqref{eq:range2} are stored in a 4$\times$4 covariance matrix ($\Sigma$-matrix).\ The square-root of the elements $\Sigma_{22}$ and $\Sigma_{44}$ represents the horizontal and vertical divergence, respectively.\ Starting with a round source beam in the transverse plane, an equal evolution of phase space in both transverse planes is expected also after the degrader.\ Therefore, the transverse divergence is defined as the average value between the divergences in the horizontal and vertical plane.\ The results of this analysis from the four MC codes are shown in \Figref{fig:phasespace}.

The slightly different corrections of Moliere's theory implemented in the fully integrated MC codes explain the discrepancies (up to maximum of 17\%) in the transverse divergence.\ In case of OPAL, the simplified elastic scattering theory based on \cite{PDG} and \cite{Jackson} leads to a maximum discrepancy of 25\% with respect to FLUKA.

\begin{figure}[h!]
  \centering
  \includegraphics[scale=0.16, keepaspectratio=true]{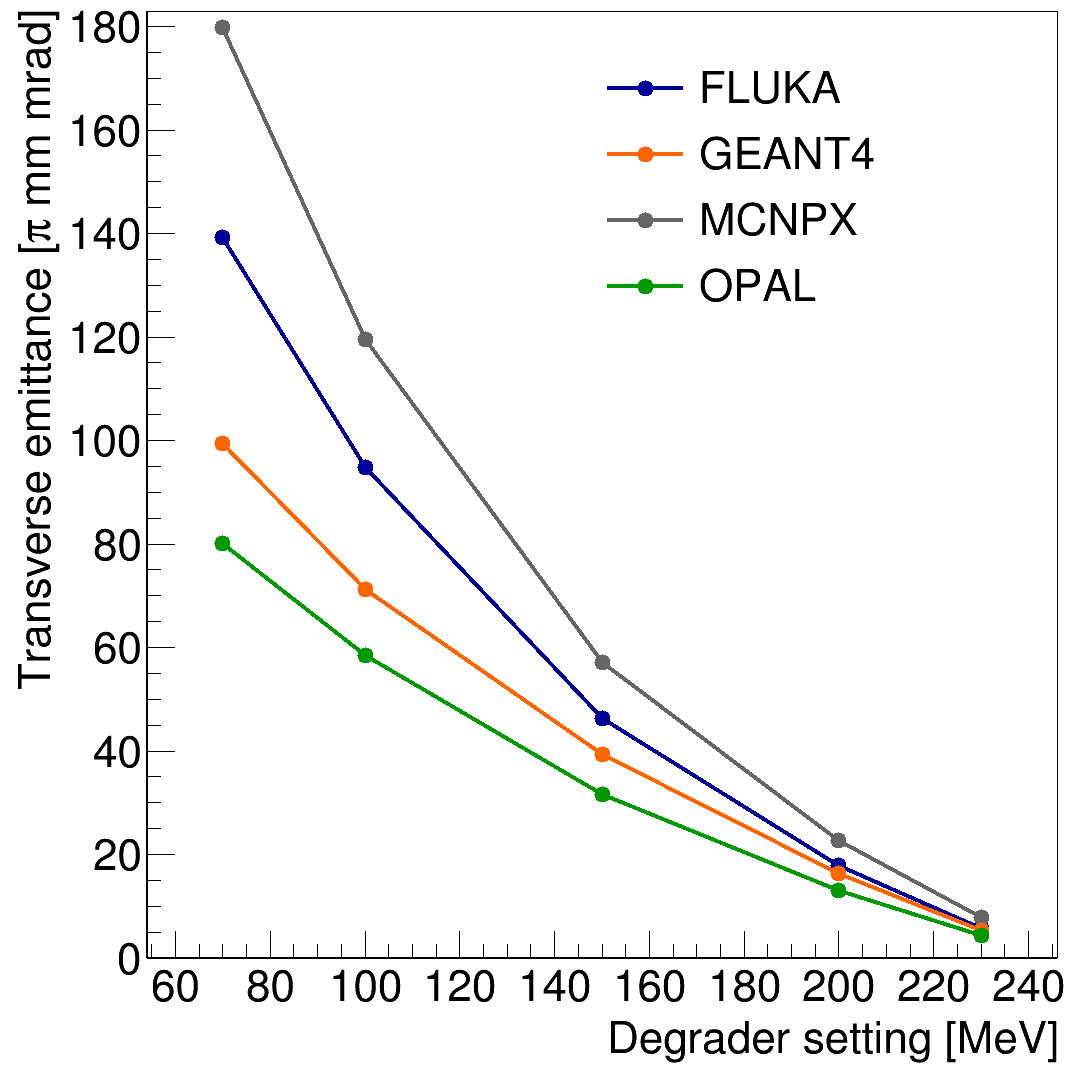}
  \caption{Growth of the transverse un-normalized rms emittance of the protons in the $\varepsilon_\text{core}$ for the different degrader settings: comparison between the four MC codes.} 
  \label{fig:emittance}
\end{figure}

The horizontal and vertical un-normalized rms emittances are evaluated from the determinant of the 2$\times$2 block of the $\Sigma$-matrix.\ Also in this case, the final transverse emittance used for the comparison between the four codes is obtained averaging the emittance in the horizontal and vertical plane.\ The emittance growth from the four MC codes is shown in \Figref{fig:emittance}.

The OPAL results are sensibly smaller than the values from the other MC codes due to the smaller divergence displayed in \Figref{fig:phasespace}.\ For MCNPX and GEANT4, the maximum discrepancy with respect to FLUKA does not exceed the 30\%.

In \secref{ssec:transmission} we discuss the impact that the differences in divergence and emittance have on the beam transmission downstream of the degrader.

\section{Application to a proton therapy facility beamline}
\label{sec:Application}

In \secref{sec:results} we analysed the main beam parameters that are normally used as starting conditions to develop a beam dynamics model.\ The comparison between the four MC codes reveal some discrepancies in the results based on the different models of the particle-matter interaction.\ In this section, we discuss the importance that the analysed parameters have in the context of a proton therapy facility and their impact on the development of a precise beam dynamics model for the transport line downstream of the degrader.

\subsection{Influence of the inelastic scattering}
\label{ssec:IS_Appl}

As shown in \Figref{fig:transmission}, excluding the inelastic scattering leads, in case of simplified codes like OPAL, to more than 30\% error in the evaluation of the transmission through the degrader.\ However, when the inelastic scattering effect is not available, the empirical formula for the total inelastic cross-section proposed in \cite{Letaw1983} can be used to estimate the number of protons that would undergo inelastic scattering.\ This empirical formula from \cite{Letaw1983} has been applied to the graphite degrader and the corresponding transmission compared with the full-physics model of FLUKA, as reported in \tabref{table:ineScatTheory}.\  

\begin{table}[ht!]
 \centering
  \caption{Transmission [\%] through the degrader from FLUKA and from the empirical formula for the inelastic scattering \cite{Letaw1983}.}
     \begin{tabular}{l|c|c}
      \hline
      \textbf{Degrader} & \textbf{FLUKA} & \textbf{Theory} \\
       \hline
	230 MeV  	& 99.16 &  94.89 \\
	200 MeV  	& 94.79 &  87.25 \\
	150 MeV  	& 84.58 &  76.89 \\
	100 MeV  	& 73.54 &  67.51 \\
	 70 MeV  	& 67.85 &  61.47 \\
        \hline
    \end{tabular}
  \label{table:ineScatTheory}
\end{table}

The theoretical values from the empirical formula (third column of \tabref{table:ineScatTheory}) can be used to scale down the constant transmission through the degrader from OPAL.\ In this way, the discrepancy between OPAL and the fully MC codes is reduced to less than 10\%, as underlined by the results in \tabref{table:ineScatTheory}.

In addition, the analysis of the inelastic scattering tail allows evaluating the losses and hence the activation of collimators or other components of the beam transport line \cite{Talanov2017, Reiss2016}.

\subsection{Beam energy and Bethe-Bloch}
\label{ssec:Ene_AP}

The kinetic energy of the proton beam after the degrader is one of the main input parameters for the beam dynamics model.\ Knowing the total graphite thickness for each setting (\tabref{table:degrpar}), the proton energy after the degrader can be calculated with the Bethe-Bloch formula from \cite{PDG}.

\tabref{table:Ene_Bethe} reports the beam energy $\bar{E}$ from the FLUKA energy spectra (with the method described in \secref{ssec:energy}) and the calculated energy values with the Bethe-Bloch equation.

\begin{table}[ht!]
 \centering
  \caption{Beam energy from the FLUKA (\tabref{table:FlukaEne}) and calculated with the Bethe-Bloch \cite{PDG}.}
     \begin{tabular}{l|c|c|c}
      \hline
      \textbf{Degrader} & \textbf{Tot. graphite}  & \textbf{FLUKA} & \textbf{Bethe-Bloch} \\
      \textbf{Setting} &  \textbf{thick. [mm]} &  \textbf{[MeV]}    & \textbf{[MeV]}\\
       \hline
	230 MeV  & 25.499 & 232.51 & 232.92  \\
	200 MeV  & 66.718 & 202.94 & 203.56 \\
	150 MeV  & 127.241 & 153.74 & 154.52 \\
	100 MeV  & 175.965 & 104.16 & 105.26 \\
	 70 MeV  & 197.190 & 76.06  & 77.38 \\
        \hline
    \end{tabular}
  \label{table:Ene_Bethe}
\end{table}

\begin{figure}[h!]
 \centering
 \includegraphics[scale=0.11, keepaspectratio=true]{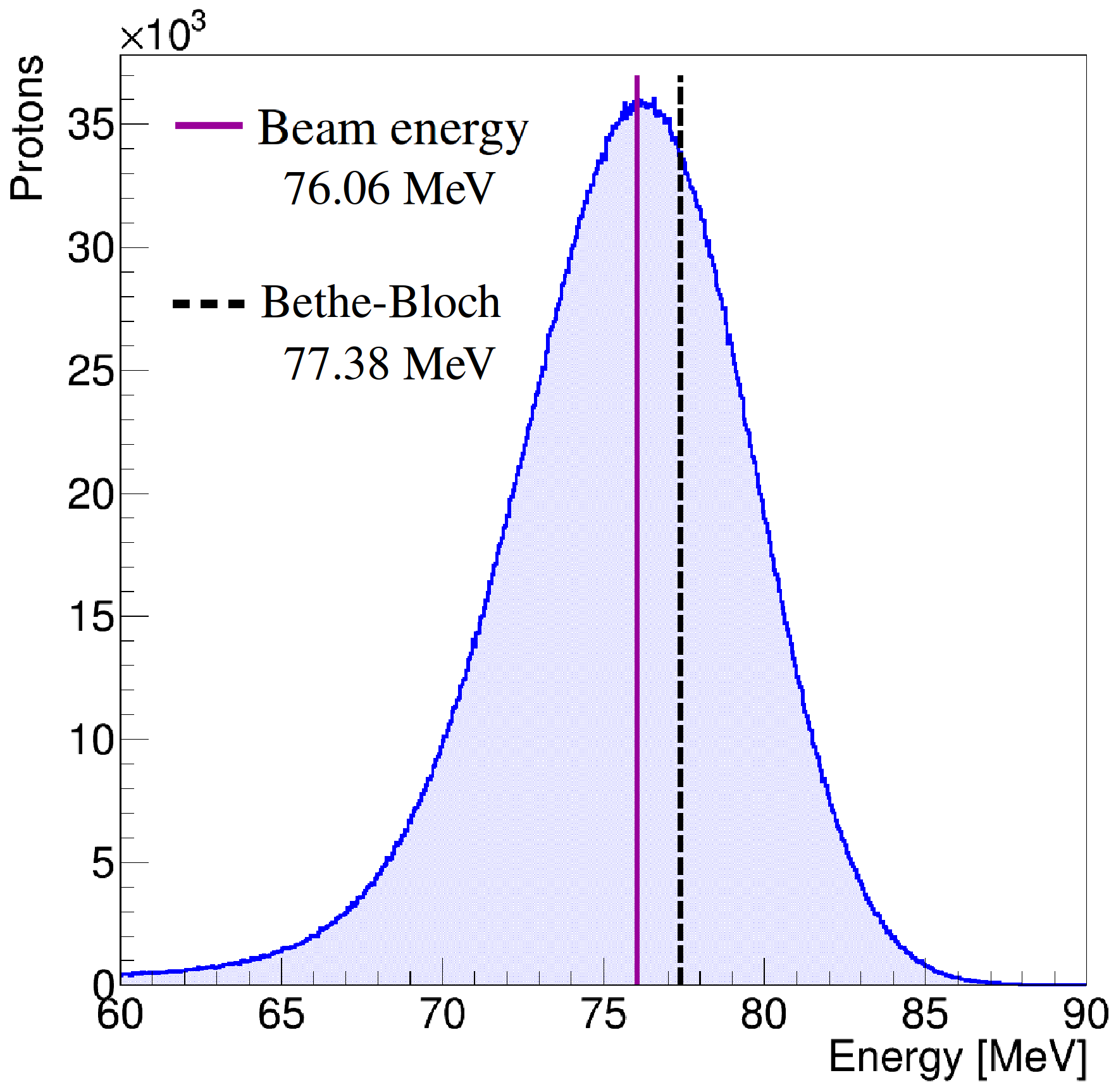}
 \caption{Comparison between the beam energy $\bar{E}$ from FLUKA and the calculated value from the Bethe-Bloch.} 
  \label{fig:Ene_Bethe}
\end{figure}

Increasing the thickness of the graphite, the energy straggling contributes to define a distribution in energy shifting down the mean energy loss with respect to the Bethe-Bloch calculation \cite{Leo, PDG}.\ An example is shown in \Figref{fig:Ene_Bethe} for 70~MeV degrader setting.\

This situation has to be taken into account when setting the elements of the ESS.\ In fact, the magnetic field of the bending magnets in the ESS are set such that the mode of the energy distribution is aligned with the aperture of the horizontal slit.\ This defines the reference energy of the beamline after the ESS that, therefore, corresponds to the beam energy $\bar{E}$ rather than to the value calculated with the Bethe-Bloch.\ The way of setting the ESS justifies our analysis of the beam energy (\secref{ssec:energy}) without reducing validity of the Bethe-Bloch calculation.

The aperture of the horizontal slit in the ESS accepts only the energies within $\pm$2\% with respect to $\bar{E}$.\ This ensure that an almost monochromatic beam is transported toward the patient.\ The energy spread values from the four MC codes in \Figref{fig:EnergySpreadComp} allow satisfying this requirement.

\subsection{Beamline transmission}
\label{ssec:transmission}

The evaluation of the transmission along the beamline downstream of the degrader is a very crucial parameter in the development of a precise beam dynamics model.

\begin{figure}[h!]
 \centering
 \includegraphics[scale=0.10, keepaspectratio=true]{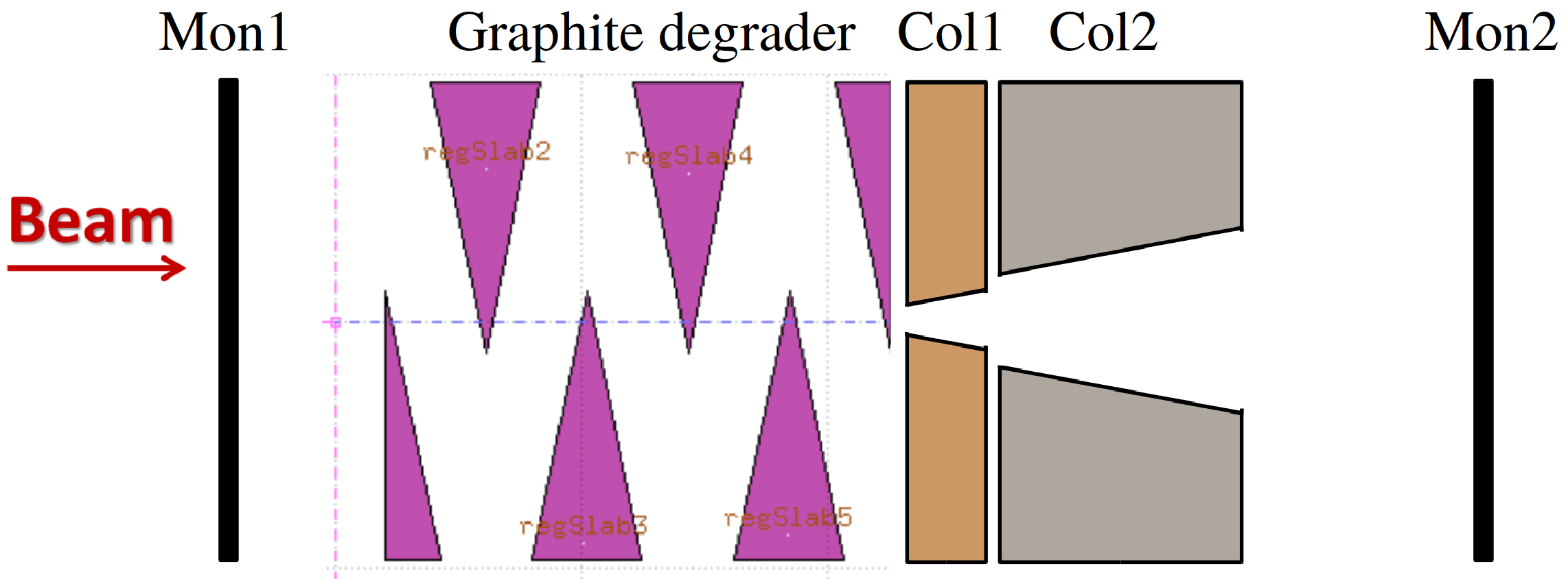}
 \caption{Schematic layout of degrader and collimators configuration in the PROSCAN facility.} 
  \label{fig:Lattice}
\end{figure}

As already mentioned, the growth of emittance and phase space is limited by collimators installed downstream of the degrader.\ A schematic layout of the degrader and collimators configuration in the PROSCAN facility is sketched in \Figref{fig:Lattice}.\ The degrader is followed by two collimators:\ the first (Col1) in copper is used to define the beam size and the second (Col2) in graphite to limit the scattered particles from Col1.\ Two strip monitors installed before the degrader (Mon1) and after Col2 (Mon2) are used to measure beam size and current during the normal operation of the facility \cite{Schippers2007}.

The accuracy of the MC models on the growth of the transverse phase space impacts the correct evaluation of the beam transmission after the collimators.\ In fact, higher losses at the collimators and hence lower transmission would be observed if the beam emerges from the degrader with a bigger emittance.\ The opposite situation with a not sufficient growth in the emittance would result in a higher transmission.

Using the typical settings of Col1 and Col2 at the PROSCAN facility, we compute the effect that these two collimators have on the scattered beam simulated from the four MC codes.\ For five degrader settings, a cut on the particle coordinates and divergences recorded at the detector plane has been applied to obtain the fraction of the scattered beam that passes the collimation system.\ The method used for this analysis is similar to the one discussed in \secref{ssec:phasespace} with \eqref{eq:range2}.\ In this case, we apply a cut of $\pm$3.5~mm (radius of Col1) on the spatial coordinates and of $\pm$21.8~mrad (angular acceptance of Col1) on the divergences.\ The second collimator (Col2) has a bigger aperture than Col1 and, therefore, does not influence this analysis.

\begin{figure}[h!]
 \centering
 \includegraphics[scale=0.18, keepaspectratio=true]{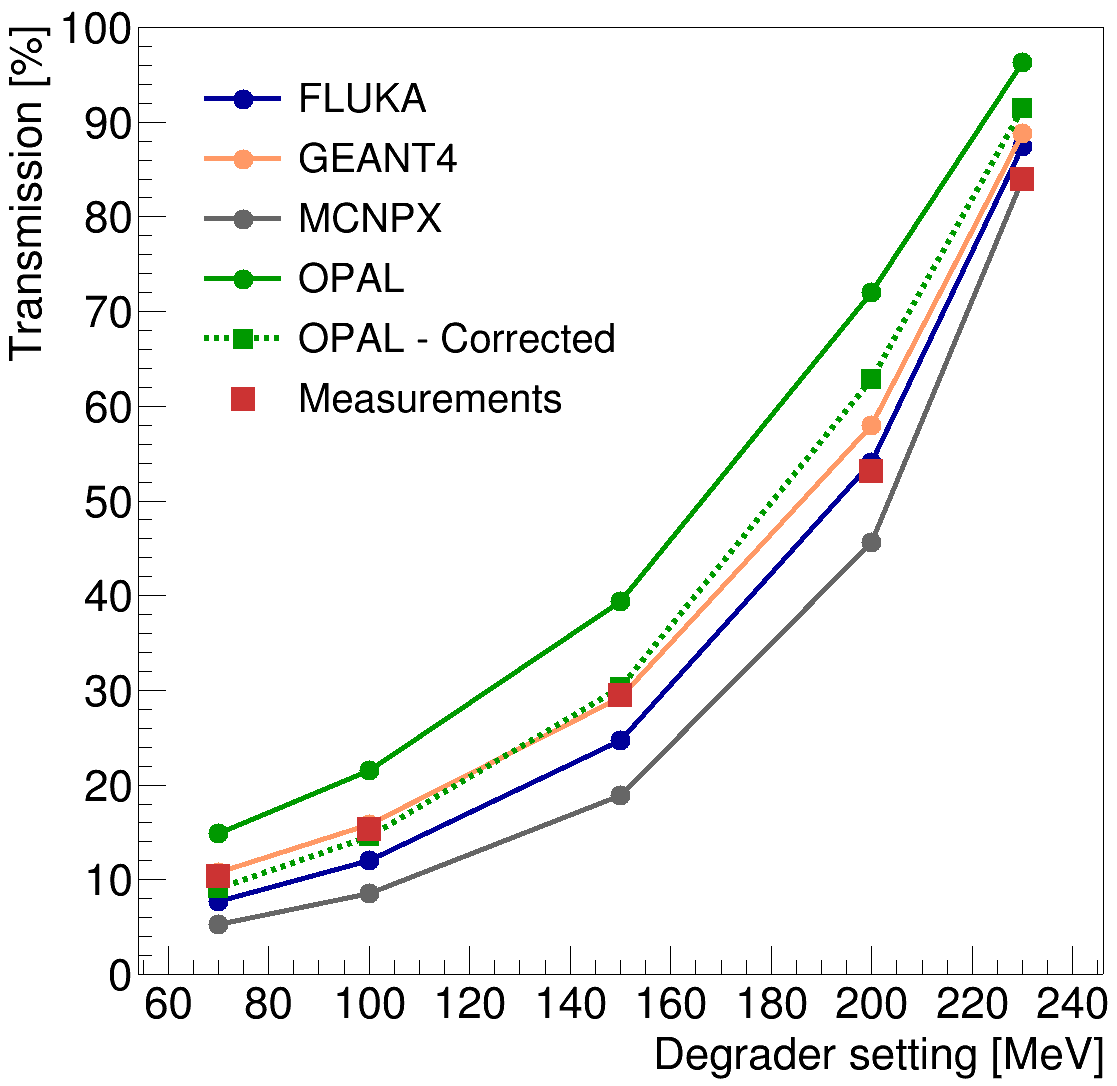}
 \caption{Beam transmission after the collimators:\ comparison between the measurement data and the results from the four MC codes.\ Two types of OPAL results are displayed:\ without inelastic scattering correction (green dots) and corrected with empirical formula for the inelastic scattering (green squares) as discussed in \secref{ssec:IS_Appl} \cite{Letaw1983}.} 
  \label{fig:Measure}
\end{figure}

Depending on the particle-matter interaction model, each MC code provides different values of the beam transmitted through the collimators.\ \Figref{fig:Measure} displays the resulting transmission from the four MC codes in comparison with measurement data performed along the PROSCAN beamline (see \Figref{fig:Lattice}).\ Using Mon1 and Mon2, the transmission after degrader and collimators is obtained by the ratio between the beam current measured at Mon2 with the nominal current before the degrader from Mon1 \cite{Baumgarten2015}.

\Figref{fig:Measure} shows that none of the codes matches the measurements in the entire energy range.\ GEANT4 results agree with the measured data in the low-middle energy range, while FLUKA and MCNPX in the higher energy range.\ The constant transmission through the degrader found in OPAL (\Figref{fig:transmission}) shows its limitation in comparison with the measurements (green dots in \Figref{fig:Measure}).\ A sensible improvement in the OPAL results is achieved when the original data have been corrected with the empirical formula for the inelastic scattering (green squares in \Figref{fig:Measure}).

\section{Conclusions}
\label{sec:conclusions}

In the context of proton therapy, a precise particle-matter interaction  model for degrader, collimators \cite{VanGoethem2009} and nozzle \cite{Paganetti2004} contributes to a better understanding of the performance of a facility in operation such as PROSCAN at the Paul Scherrer Institut \cite{PROSCAN}, as well as to support the design and optimisation of new proton therapy solutions.\ In particular, the results from the degrader model in this work are contributing to the design of a new superconducting gantry prototype at PSI where the degrader is mounted on the gantry itself \cite{Talanov2017, SCGantry}.

In this work we compared the results of four different MC codes used to simulate the effects of a graphite degrader.\ The implementations of the particle-matter interaction model of each code led to different outcomes on the degrader model and beam properties, as discussed in \secref{sec:results}.\ In \secref{sec:Application} we discussed the outcomes from the four MC codes in the context of a proton therapy facility.\ For such applications, the use of fully integrated MC codes is of course advantageous in terms of precision and accuracy of the outcomes.\ If the degrader model is extended also to the downstream transport line, then a combination with specific beam dynamics codes for the optics simulations is required.\ In such a case, the use of simplified particle-matter interaction model or single code like OPAL that combines particle tracking and MC capabilities is often preferred with respect to the fully integrated MC codes \cite{Rizzoglio2017}.

From our analysis, we can conclude that in a cyclotron-based proton therapy facility the presence of collimators and ESS allows reducing the discrepancies between fully integrated MC codes and a simplified particle-matter interaction model.\ In fact, collimators suppress the inelastic scattering tail and limit the elastic Multiple Coulomb scattering that contributes to the phase space growth.\ We have shown the importance to apply corrections (when possible) to compensate the lack of accuracy in the simplified particle-matter interaction model.\ This would lead to a better agreement of the model in comparison against measurements. 


\section*{Acknowledgements}
The authors wish to thank Thomas Schietinger for the proofreading of the manuscript.


\section*{References}
\bibliography{MonteCarloCodes.bib}

\begin{thebibliography}{10}

\bibitem{HPaganetti}
H.~Paganetti.
\newblock {\em {Proton Therapy Physics}}.
\newblock {CRC Press, Taylor \& Francis Group}, 2011.

\bibitem{Anferov2007}
V.~Anferov \textit{et al.}
\newblock {Indiana University Cyclotron Operation for Proton Therapy Facility}.
\newblock pages (231--233), 2007.
\newblock {Cyclotron Conference}.

\bibitem{Gerbershagen2016}
A.~Gerbershagen \textit{et al.}
\newblock {Measurements and simulations of boron carbide as degrader material
  for proton therapy}.
\newblock {\em Phys. Med. Biol.}, 61(14):337--348, 2016.

\bibitem{Anferov2003}
V.~Anferov.
\newblock {Energy degrader optimization for medical beam lines}.
\newblock {\em Nucl. Instr and Meth. A}, 496:222--227, 2003.

\bibitem{Lomax}
C.-M.~Charlie Ma and Tony Lomax.
\newblock {\em {Proton and Carbon Ion Therapy}}.
\newblock {CRC Press, Taylor \& Francis Group}, 2013.

\bibitem{Fluka}
A.~Ferrari, P.R. Sala, A.~Fass{\'o}, and J.~Ranft.
\newblock {\em {FLUKA: a multi-particle transport code}}, 2005.
\newblock CERN Yellow report, CERN-2005-10.

\bibitem{Geant4}
S.~Agostinelli \textit{et al.}
\newblock {Geant4 - A simulation toolkit}.
\newblock {\em Nucl. Instr and Meth. A}, 506:250--303, 2003.

\bibitem{MCNPX}
D.B. Pelowitz.
\newblock {\em {MCNPX Users Manual Version 2.7.0}}, 2011.
\newblock LA-CP-11-00438.

\bibitem{Gottschalk2010}
B.~Gottschalk.
\newblock {On the scattering power of radiotherapy protons}.
\newblock {\em Med. Phys}, 37(1):352--367, 2010.

\bibitem{Schneider2001}
U.~Schneider \textit{et al.}
\newblock {On small angle multiple Coulomb scattering of protons in the
  Gaussian approximation}.
\newblock {\em Z. Med. Phys.}, 11:110--118, 2001.

\bibitem{Farley2005}
J.~M. Farley.
\newblock {Optimum strategy for energy degraders and ionization cooling}.
\newblock {\em Nucl.Inst.Meth. A}, 540(2-3):235--244, 2005.

\bibitem{VanGoethem2009}
M.J. van Goethem~\textit{et al.}
\newblock {Geant4 simulations of proton beam transport through a carbon or
  beryllium degrader and following a beam line}.
\newblock {\em Phys. Med. Biol.}, 54(19):5831--46, 2009.

\bibitem{opal:1}
A.~Adelmann, C.~Baumgarten, M.~Frey, A.~Gsell, V.~Rizzoglio, J.~Snuverink
  (PSI), C.~Metzger-Kraus, Y.~Ineichen, X.~Pang, C.~Wang S.~Russell~(LANL),
  J.~Yang (CIAE), S.~Sheehy, Chris~Rogers (RAL), and D.~Winklehner (MIT).
\newblock {The OPAL (Object Oriented Parallel Accelerator Library) Framework}.
\newblock Technical Report PSI-PR-08-02, Paul Scherrer Institut, 2008 - 2017.

\bibitem{Rizzoglio2017}
V.~Rizzoglio, A.~Adelmann, C.~Baumgarten, M.~Frey, A.~Gerbershagen, D.~Meer,
  and J.M. Schippers.
\newblock {Evolution of a beam dynamics model for the transport line in a
  proton therapy facility}.
\newblock {\em Phys. Rev. Accel. Beams}, 20:124702--124714, 2017.

\bibitem{PROSCAN}
http://www.psi.ch/protontherapy.

\bibitem{Reist2002}
H.~Reist \textit{et al.}
\newblock {A fast degrader to set the energies for the application of the depth
  dose in proton therapy}.
\newblock Technical Report Volume V, Paul Scherrer Institut, 2002.

\bibitem{Battistoni2016}
G.~Battistoni \textit{et al.}
\newblock {The FLUKA Code: An Accurate Simulation Tool for Particle Therapy}.
\newblock {\em Front. Oncol}, 6(116):1--24, 2016.

\bibitem{Ferrari1992}
A.~Ferrari, P.R. Sala, R.~Guaraldi, and F.~Padoani.
\newblock {An improved multiple scattering model for charged particle
  transport}.
\newblock {\em Nucl.Inst.Meth. B}, 71(4):412--426, 1992.

\bibitem{BDSIM}
L.~J.~Nevay \textit{et al.}
\newblock {Beam Delivery Simulation: {BDSIM} automatic Geant4 models of
  accelerators}.
\newblock {\em Proceedings of IPAC2016}, pages 3098--3100, 2016.

\bibitem{Lewis1950}
H.~W. Lewis.
\newblock {Multiple Scattering in an Infinite Medium}.
\newblock {\em Phys. Rev.}, 78:526--529, 1950.

\bibitem{Stachel2013}
H.~Stachel.
\newblock {Double Degrader for Proton Therapy}.
\newblock Master's thesis, ETH Z{\"u}rich, 2013.
\newblock
  http://amas.web.psi.ch/people/aadelmann/ETH-Accel-Lecture-1/projectscompleted/phys/stachel.pdf.

\bibitem{Andersen}
H.~H. Andersen and J.~F. Ziegler.
\newblock {\em {Hydrogen Stopping Power and Ranges in all Elements}}.
\newblock Pergamon, New York, 1977.

\bibitem{PDG}
K.A.~Olive \textit{et al.}
\newblock {\em Particle Data Group}.
\newblock Chin. Phys. C, 2014.

\bibitem{Jackson}
J.D. Jackson.
\newblock {\em Classical Electrodynamics}.
\newblock John Wiley and Sons Ltd., third edition edition, 1962.

\bibitem{Letaw1983}
J.R. Letaw, R.~Silberberg, and C.H. Tsao.
\newblock {Proton-nucleus total inelastic cross sections: an empirical formula
  for E $>$ 10 MeV}.
\newblock {\em The Astrophysical Journal Supplement Series}, 51:271--276, 1983.

\bibitem{Talanov2017}
{V. Talanov}, {D. Kiselev}, {D. Meer}, {V. Rizzoglio}, {J.M. Schippers}, {M.
  Seidel}, and {M. Wohlmuther}.
\newblock {Neutron doses due to beam losses in a novel concept of a proton
  therapy gantry}.
\newblock {\em {Journal of Physics: Conference Series}}, 874(1):1--5, 2017.

\bibitem{Reiss2016}
{T. Reiss}, {V. Talanov}, {M. Wohlmuther}, {D. Kiselev}, {R. Scheibl}, {D.
  Mohr}, and {A. Fuchs}.
\newblock {Shielding calculations for Gantry 3}.
\newblock Technical Report TM-8-14-01 Rev.1, Paul Scherrer Institut, 2016.

\bibitem{Leo}
W.R. Leo.
\newblock {\em Techniques For Nuclear And Particle Physics Experiments}.
\newblock Springer-Verlag, second revised edition edition, 1994.

\bibitem{Schippers2007}
J.M.~Schippers \textit{et al.}
\newblock {The SC cyclotron and beam lines of PSI's new protontherapy facility
  PROSCAN}.
\newblock {\em Nucl. Instr and Meth. B}, 261(1):773--776, 2007.

\bibitem{Baumgarten2015}
C.~Baumgarten.
\newblock {Gantry 3 Transmission}.
\newblock Technical report, Paul Scherrer Institut, 2015.

\bibitem{Paganetti2004}
H.~Paganetti, H.~Jiang, S.Y. Lee, and H.M. Kooy.
\newblock {Accurate Monte Carlo simulations for nozzle design, commissioning
  and quality assurance for a proton radiation therapy facility}.
\newblock {\em Med. Phys.}, 31(7):2107--18, 2004.

\bibitem{SCGantry}
A.~Gerbershagen \textit{et al.}
\newblock {A novel beam optics concept in a particle therapy gantry utilizing
  the advantages of superconducting magnets}.
\newblock {\em Z. Med. Phys.}, 26:224--237, 2016.

\end{thebibliography}
\bibliographystyle{unsrt}

\end{document}